\documentclass[a4paper,11pt]{article}

\usepackage{jheppub}
\usepackage{mathrsfs}
\usepackage{verbatim}
\usepackage{booktabs}
\usepackage[tight]{subfigure}

\newcommand{\be}{\begin{equation}}
\newcommand{\ee}{\end{equation}}
\newcommand{\bea}{\begin{eqnarray}}
\newcommand{\eea}{\end{eqnarray}}

\newcommand{\df}{\dfrac}

\def\simlt{\stackrel{<}{{}_\sim}}
\def\simgt{\stackrel{>}{{}_\sim}}
\newcommand{\gsim}{\lower.7ex\hbox{$\;\stackrel{\textstyle>}{\sim}\;$}}
\newcommand{\lsim}{\lower.7ex\hbox{$\;\stackrel{\textstyle<}{\sim}\;$}}
\newcommand{\A}{\mathscr{A}}

\title{ Implications of light charginos for Higgs observables, LHC searches and dark matter}
\author{J. Alberto Casas,} \author{Jes\'us M. Moreno,} \author{Krzysztof Rolbiecki} \author{and Bryan Zald\'ivar}
\affiliation{Instituto de F\'isica Te\'orica, IFT-UAM/CSIC \\
        U.A.M., Cantoblanco, 
        28049 Madrid, Spain }
\emailAdd{alberto.casas@uam.es} \emailAdd{jesus.moreno@csic.es} \emailAdd{krzysztof.rolbiecki@desy.de} \emailAdd{b.zaldivar.m@csic.es} 

\abstract{
A rather high Higgs mass, $m_h\simeq 126$ GeV, suggests that at least a part of the supersymmetric spectrum of the MSSM may live beyond $\mathcal{O}(1~\mathrm{TeV})$ and hence inaccessible to the LHC. However, there are theoretical and phenomenological reasons supporting a possibility that charginos and neutralinos remain much closer to the electroweak scale. In this paper, we explore such a scenario in the light of recent Higgs measurements, mainly its di-photon decay rate, where the data might indicate a slight excess over the SM prediction. That excess could be fitted by the contribution of light charginos provided $\tan\beta$ is low to moderate, a possibility that is receiving much attention for other theoretical reasons. We investigate the implications of this scenario for other observables, such as dark matter constraints, electroweak observables and experimental signals at the LHC, like  di-lepton, tri-lepton and same-sign dilepton. An important part of the models survive all the constraints and are able to give positive signals at LHC-14TeV and/or XENON1T.}

\preprint{IFT-UAM/CSIC-13-060}

\begin{document}
\maketitle
\flushbottom

\section{Introduction}

The evidence obtained by ATLAS~\cite{ATLAS:2013mma} and CMS~\cite{CMS:yva} of the Standard Model (SM)-like Higgs boson with mass $m_h\simeq 126$ GeV has deep implications for supersymmetry (SUSY). As it is well known, such a value is causing a tension with naturalness within the minimal supersymmetric standard model (MSSM) since it requires the stop sector to be quite heavy on average (typically well above 1 TeV unless the stop mixing is close to the maximal value), thus suggesting that the mass scale of SUSY particles could be substantially higher than expected from fine-tuning arguments. Actually, if {\em all} the supersymmetric particles have masses of the same order as the stop average, then SUSY is essentially decoupled from all the low-energy observables, including Higgs production and decay. In that unfortunate instance, it will be very challenging to detect such SUSY at the LHC. In fact, this is already the most likely situation for the constrained MSSM~\cite{Farina:2011bh,Balazs:2012qc,Akula:2012kk,Buchmueller:2012hv,Arbey:2012dq,Strege:2012bt,Cabrera:2012vu}, i.e.\ assuming universality of soft terms at a high-scale.

The prospects become much more interesting if some supersymmetric states remain sufficiently light. This is not just wishful thinking, there are theoretical and phenomenological arguments pointing out to this possibility. First of all, it is highly desirable that the fermionic partners of the SM particles are around the TeV range in order to keep the successful gauge unification that occurs in the MSSM. Besides, one of those fermionic states, the lightest neutralino, is an excellent candidate for Dark Matter (DM), provided it is not too far from the weak scale\footnote{For an early analysis in this context, see \cite{Belanger:2000tg}.}. These arguments go in the direction of a split-SUSY-like scenario~\cite{ArkaniHamed:2004fb,Giudice:2004tc}, though the scalar particles do not need to live at an extremely high scale. Incidentally, a spectrum with heavy sfermions avoids potential flavour violation problems. In a scenario of this kind one can keep the hope of detecting SUSY either in a direct way (production of neutralinos, charginos and maybe gluinos) and/or in an indirect one, by measuring deviations of Higgs couplings from SM values. In the latter case, it is remarkable that all the current production and decay data about the Higgs boson are so far consistent with the SM values~\cite{ATLAS:2013mma,CMS:yva}, although the uncertainties are still large and some of the measurements are not centered at the corresponding SM prediction. So there is still room for non-decoupled new-physics to be unveiled in the future. In this spirit, probably the most interesting observable is still the Higgs decay into two photons.

Last reported analyses of ATLAS and CMS on $h\to\gamma\gamma$ Higgs decay events, though consistent with each other within $2\sigma$, show a different level of agreement with the Standard Model (SM). While ATLAS continues to observe a $\sim 2\sigma$ excess over the SM prediction~\cite{ATLAS-CONF-2013-012}, CMS has become consistent with it within $1\sigma$~\cite{CMS-PAS-HIG-13-001}:
\bea
{\rm ATLAS:}\hspace{1cm} &\sigma/\sigma_{SM}&=1.65^{+0.34}_{-0.30}\,,
\nonumber\\
{\rm CMS^{(1)}:}\hspace{1cm} &\sigma/\sigma_{SM}&=0.78^{+0.28}_{-0.26}\hspace{0.3cm}{\rm (MVA\ mass-factorized)},
\nonumber\\
{\rm CMS^{(2)}:}\hspace{1cm} &\sigma/\sigma_{SM}&=1.11^{+0.32}_{-0.30}\hspace{0.3cm}{\rm (Cut-based)}.
\eea
One can attempt to combine these results using the principle of maximum likelihood as ``best estimator". Then,
\bea
{\rm ATLAS-CMS^{(1)}:}\hspace{1cm} &\sigma/\sigma_{SM}&=1.14\pm 0.20\,,\hspace{3cm}
\nonumber\\
{\rm ATLAS-CMS^{(2)}:}\hspace{1cm} &\sigma/\sigma_{SM}&=1.37\pm 0.22\,,\hspace{3cm}
\label{ATLASCMS2}
\eea
where we have averaged the positive and the negative uncertainties to perform the combination.

Obviously we cannot know the future evolution of the ATLAS and CMS central values as statistics increases. It is very likely that both, and thus the combined result, will continue converging towards the SM. But if the excess persists, with smaller uncertainties, it would become a possible indication that the Higgs boson is not exactly the SM one. In that case (non-decoupled) SUSY would be a most natural framework to explain the discrepancies.

Since, up to now, the Higgs production rate and the Higgs decay in other channels seem to be quite consistent with the SM expectations, the new contributions should essentially affect only $h\to\gamma\gamma$ (and maybe other not-yet well measured decay modes). Then within SUSY, the natural sources of the $h\to\gamma\gamma$ enhancement arise from one-loop contributions involving charged Higgses, staus or charginos. These possibilities have been considered in the literature long ago, see e.g.\ refs.~\cite{Gunion:1989we,Diaz:2004qt,Carena:2012gp,Djouadi:2005gi,Djouadi:2005gj,Arganda:2013ve}. 

Obviously, in the context of heavy sfermions, the chargino contribution appears as the only viable one.\footnote{
The possibility of a substantial stau-contribution was considered in detail in ref.~\cite{Carena:2012gp}, where it was shown that one can get a considerable excess in $h\to\gamma\gamma$ provided the staus are close to their present LEP lower limit (85-90 GeV) and $\tan\beta$ is very large. In addition, such a scenario is consistent with the observed DM relic density, provided the LSP neutralino is a few tens of GeV lighter than the light stau.} The chargino enhancement of $h\to\gamma\gamma$ has been reported to be $< 10\%$ for usual values of the supersymmetric parameters (this was too small for the previous excess reported by ATLAS and CMS~\cite{Aad:2012tfa,Chatrchyan:2012ufa}), unless one goes to extended scenarios beyond the   MSSM~\cite{Huo:2012tw}. However,  the chargino contribution incresases for lighter charginos and low $\tan\beta$. As we will see, there are narrow regions of the MSSM parameter-space where the enhancement in the $h\to\gamma\gamma$ rate is up to 25\%, and wider 
regions leading to a substantial increase around 10--20\%. Hence, if the $h\to\gamma\gamma$ excess persists in the future at values similar to those of eq.~(\ref{ATLASCMS2}), with reduced uncertainty, the chargino contribution might become a very natural way to explain it. On the other hand, the DM implications of such a scenario have not yet been explored, as well as its possible LHC signals. 

The main goal of this paper is to study these items in detail, i.e.\ to explore to which extent the contribution of charginos to $h\to\gamma\gamma$ can enhance substantially the signal and if the corresponding parameter space is consistent with other experimental bounds coming from DM, e.g.\ PLANCK~\cite{Ade:2013uln} measurement of the relic density and XENON~\cite{Lavina:2013zxa} bound on direct detection, as well as with present bounds on the invisible decay width of the Higgs~\cite{Falkowski:2013dza,Giardino:2013bma,Belanger:2013kya,Djouadi:2013qya, Espinosa:2012vu} and with electroweak observables. We will also investigate the possible signals of this kind of scenario at the LHC and thus the 
present constraints and the prospects of being tested in the future. Furthermore we discuss the opportunities of this framework to be discarded or discovered by experiments of direct DM detection, as XENON1T.

Interestingly, scenarios of low $\tan\beta$ used to be considered as disfavoured by naturalness arguments. Since the tree level Higgs mass goes like $m_h^2\sim M_Z^2 \cos^2 2\beta$, an ${\cal O}(1)$ value of $\tan\beta$ requires large radiative corrections to reproduce the Higgs mass above 114 GeV. Such large radiative corrections require at least one heavy stop, which in turn communicates to the Higgs mass-parameters via RG equations, making the EW breaking process fine-tuned. However, the actual Higgs mass, $m_h\simeq 126$ GeV, is so high that essentially any MSSM scenario will require heavy stops and thus fine-tuning. This fact has re-opened the small $\tan\beta$ regime as an interesting one, since possibly it is not qualitatively worse than the large $\tan\beta$ one; see the recent discussion in ref.~\cite{Djouadi:2013vqa}. Actually, although the low $\tan\beta$ regime certainly requires heavier stops (for the extreme case $\tan\beta=1$ their masses should lie at $10^{6-10}$ GeV), on the other hand it presents some attractive features. As shown in ref.~\cite{Ibanez:2013gf},  ${\cal O}(1)$ $\tan\beta$ values are favoured when the supersymmetric particles are very heavy, amusingly because they amount to much less fine-tuning. The string-theoretic motivations for such low $\tan\beta$ have been recently discussed in ref.~\cite{Hebecker:2013lha}. Clearly, the fact that this scenario can provide a positive signal in $h\to\gamma\gamma$ adds interest to this possibility.

In section~\ref{sec:2} we recapitulate SM and SUSY contributions to $h\to \gamma \gamma$ decay and study in detail the chargino impact. In section~\ref{sec:3} we introduce representative benchmark models giving a sizeable contribution to the Higgs decay, consistent with LEP limits. Section~\ref{sec:4} is devoted to the analysis of dark matter bounds in each of the benchmark scenarios. In section~\ref{sec:STU} we evaluate the predictions for the $S, T, U$ parameters. Finally, in section~\ref{sec:constraints} we study direct LHC searches of SUSY-EW particles and conclude in section~\ref{sec:conclusions}.

\section{Charginos and $h\to\gamma\gamma$\label{sec:2}}
\subsection{$h\to\gamma\gamma$ in the SM}

The formulation of the SM contribution to  $h\to\gamma\gamma$ can be found e.g.\ in ref.~\cite{Djouadi:2005gi}. It comes essentially from $W$-bosons and heavy fermions running inside the loop through which the Higgs decays. The expression for the decay rate is:
\be
\Gamma(h\to\gamma\gamma)=\df{G_\mu}{128}\df{\alpha^2 m_h^3}{\sqrt{2}\pi^3}
\left| \sum_f N_c Q_f^2 \mathscr{A}_{1/2}^h (x_f) + \A_1^h (x_W)\right|^2 \equiv \df{G_\mu}{128}\df{\alpha^2 m_h^3}{\sqrt{2}\pi^3} |A^h_{SM} |^2~,
\label{h2AAsm}
\ee
where the form-factors of spin-1/2 and spin-1 particles are given by
\bea
\A_{1/2}^h (x) &=& 2[x + (x-1) f(x)] x^{-2}~,\nonumber \\
\A_1^h (x) &=& -[2x^2 + 3x  + 3(2x-1) f(x)]x^{-2} \,,
\eea
with the loop functions defined as
\begin{displaymath}
f(x) = \left\{ 
\begin{array}{cc}
\arcsin^2\sqrt{x} & ~~~x\leq 1 \\
-\df{1}{4}\left[\log \df{1+\sqrt{1-x^{-1}}}{1-\sqrt{1-x^{-1}}} - i\pi \right]^2 &~~~ x >1
\end{array}
\right.
\end{displaymath}
and $x_i \equiv m_h^2/4M_i^2$, with $i=f,W$.

\subsection{Supersymmetric contributions to $h\to\gamma\gamma$}

\begin{figure}[t]
\centering
\includegraphics[width=0.3\textwidth,angle=0]{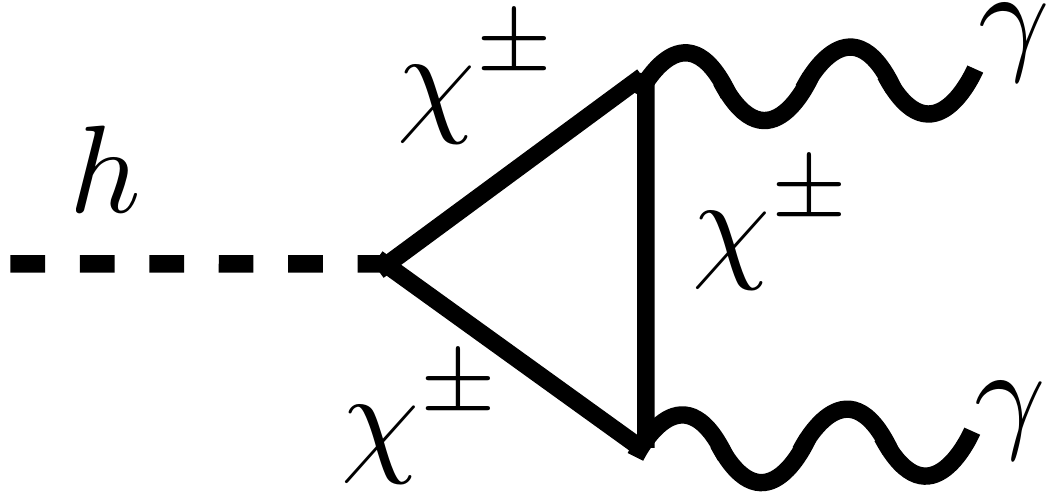}
\caption{ Feynman diagram of chargino one-loop contribution to the $h\to\gamma\gamma$ process. }
\label{fig:h2aa}
\end{figure}

In SUSY, the previous SM expression for $\Gamma(h\to\gamma\gamma)$ is slightly altered due to modifications in the couplings of the lightest Higgs to the SM particles. In addition there appear new supersymmetric contributions from decay diagrams involving charged Higgses, charged sfermions or charginos. We use here the formulation expounded in ref.~\cite{Djouadi:2005gj}, namely:
\bea
\label{h2AAsusy}
\Gamma(h\to\gamma\gamma)&=&\df{G_\mu}{128}\df{\alpha^2 m_h^3}{\sqrt{2}\pi^3}
\left| \sum_f N_c Q_f^2 g_{hff} \A_{1/2}^h (x_f) + g_{hWW} \A_1^h (x_W) \right. \\
&+& \left. \df{M_W^2 \lambda_{hH^+H^-}}{2c_W^2 M^2_{H^\pm}} \A_0^h (x_{H^\pm}) + A^h_{\rm SUSY} 
\right|^2 
\nonumber
\eea
where the SM-like couplings in eq.~(\ref{h2AAsm}) are now modified by additional prefactors $g_{hff}$, $g_{hWW}$,
\bea
g_{h\bar uu} = \cos\alpha/\sin\beta,~~~~g_{h\bar dd} = -\sin\alpha/\cos\beta,~~~~g_{hWW} = \sin(\beta-\alpha) \,.
\eea
Here $\alpha$ is the mixing angle of the two CP-even neutral Higgses and $\tan \beta$ is the ratio of the two Higgs VEVs, $\tan\beta = \langle H_u\rangle / \langle H_d\rangle$. In the so-called ``decoupling limit" (for the Higgs sector), where the pseudoscalar mass is large, $m_A \gg m_h$ (as happens in our analysis), one simply has $\tan\alpha \approx -1/\tan\beta$, so the SM contribution remains essentially the same. The contribution of the charged Higgses is proportional to the $\lambda_{hH^+H^-}$ coupling, 
\bea
\lambda_{hH^+H^-} &=& \cos2\beta \sin(\beta+\alpha) + 2c_W^2\sin(\beta-\alpha)~,
\eea
and to the scalar form-factor of the interaction,
\be
\A_0^h (x) = - [x - f(x)] x^{-2}~.
\ee
The rest of the SUSY contributions come from the chargino and sfermion loops and are collected in the $A^h_{\rm SUSY}$ piece,
\bea
\label{Asusy}
A^h_{\rm SUSY} &=& \sum_{\chi_i^\pm} \df{2 M_W}{m_{\chi_i^\pm}} g_{h\chi_i^+\chi_i^-} \A^h_{1/2}(x_{\chi_i^\pm})
+ \sum_{\tilde f_i} \df{g_{h\tilde f_i\tilde f_i}}{m^2_{\tilde f_i}} N_c Q^2_{\tilde f_i} \A^h_0 (x_{\tilde f_i})\nonumber\\
&\equiv& A_{\chi^\pm}^h~ + A_{\tilde{f}}^h\,.
\eea
We will focus now on the chargino contribution, which arises from the diagram shown in figure~\ref{fig:h2aa}. Note that the coupling of the Higgs to the charginos requires that the latter are a non-trivial mixing of the wino and Higgsino components. The coupling of the Higgs to the charginos  is given by: 
\bea
g_{h\chi^+_i\chi^-_i} &=& g^L_{h\chi^+_i\chi^-_i} P_L + g^R_{h\chi^+_i\chi^-_i} P_R, \nonumber \\
&=& \df{1}{\sqrt{2}} (-\sin\alpha V_{i1} U_{i2} + \cos\alpha V_{i2} U_{i1})~,
\label{ghchi}
\eea
where $U,V$ are the (unitary) chargino mixing-matrices, satisfying 
\be
U^* m_{\chi^\pm} V^\dag = {\rm diag}(m_{\chi^\pm_1},m_{\chi^\pm_2})
\ee 
and $m_{\chi^\pm}$ is the chargino mass matrix
\begin{displaymath}
 m_{\chi^\pm}= \left(
\begin{array}{cc}
M_{2} & \sqrt{2} M_W \sin\beta \\
\sqrt{2} M_W \cos\beta & \mu
\end{array}
\right) ~.
\label{X}
\end{displaymath}
For real values of $M_2$ and $\mu$, as we are assuming here, both $U$ and $V$ are orthogonal. Then $g^L_{h\chi^+_i\chi^-_i}=g^R_{h\chi^+_i\chi^-_i}$, thus the simplified expression in the second line of eq.~(\ref{ghchi}).

\begin{figure}[t]
\centering
\includegraphics[width=0.4\textwidth,angle=0]{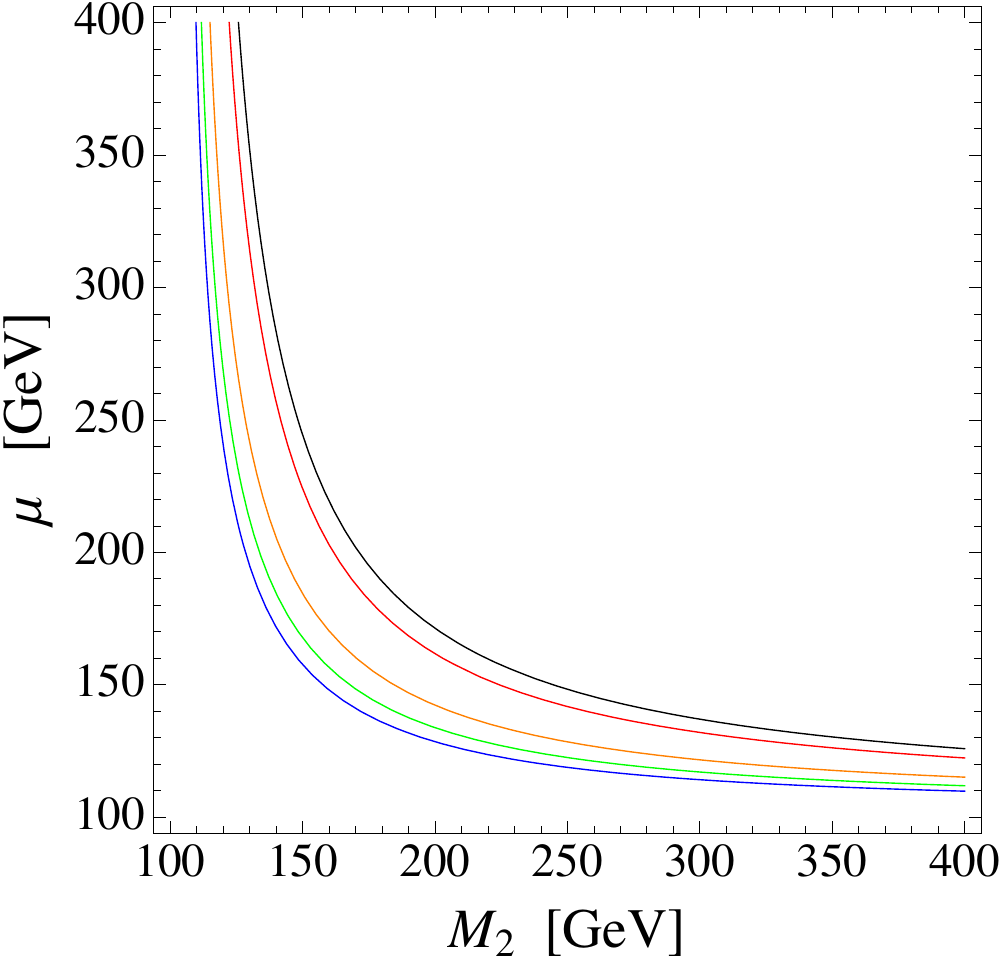}
\caption{ Contour plot in the $M_2,\mu$ plane giving $m_{\chi^\pm_1}=104$ GeV, for $\tan\beta = 1,2,5,10$ and 25, shown in black (upper line), red, orange, green and blue (lower line), respectively. For every contour, the region to the left gives smaller (thus ruled-out) values of $m_{\chi^\pm_1}$. }
\label{fig:Mchi}
\end{figure}

The parameters involved in the chargino mass matrix, i.e.\ $M_2$, $\mu$ and $\tan\beta$, are constrained by the present
LEP bound on the lightest-chargino mass, namely $m_{\chi^\pm_1}\gtrsim 104$ GeV~\cite{Beringer:1900zz}. As can be seen in figure~\ref{fig:Mchi}, the smaller the value of $\tan\beta$ the stronger the corresponding limits on $M_2$ and $\mu$.

\subsection{The size of the chargino contribution}

In the present analysis we assume that the sfermion and the charged Higgs sectors are heavy enough not to affect in an appreciable amount the supersymmetric loop contribution to $h\to\gamma\gamma$, so that the latter is dominated by the chargino contribution.  As mentioned in the Introduction, this is a quite plausible possibility since, as it is well-known, the rather large Higgs mass, $m_h\simeq 126$ GeV, requires heavy stops, $m_{\tilde t}\simgt 1$ TeV. Hence, if the MSSM under consideration incorporates universality of scalar masses, clearly all the sfermion contributions to $h\to\gamma\gamma$ are negligible. Certainly, allowing for non-universality of scalar masses opens the possibility of increasing $h\to\gamma\gamma$ substantially without modifying appreciably the Higgs production thanks to the contribution of light staus, provided $\tan\beta$ is very large \cite{Carena:2012gp}. In contrast, for a substantial chargino contribution the requirements are different: one needs a rather light chargino and moderate-to-low values of $\tan\beta$. Thus, it is unlikely that the sfermion and the chargino contributions co-operate to increase the rate of $h\to\gamma\gamma$ in a relevant way. The supersymmetric contribution is only important when either one or the other source is dominant.

In our scenario, the contribution of charginos to $h\to\gamma\gamma$, $A^h_{\chi^\pm}$, is the only relevant contribution to the $A^h_{\rm SUSY}$ term of eqs.~(\ref{h2AAsusy}) and~(\ref{Asusy}). Thus
\be
\Gamma[h\to\gamma\gamma] = \df{G_\mu}{128}\df{\alpha^2 m_h^3}{\sqrt{2}\pi^3} |A^h_{SM} + A^h_{\chi^\pm} |^2~.
\label{h2AAC}
\ee

The sign of $A^h_{\chi^\pm}$can be the same as the $A^h_{SM}$ term. This occurs in particular when both $M_2$ and $\mu$ have positive sign.\footnote{For $M_2>0, \mu < 0$ the interference between wino and Higgsino components becomes negative, leading to suppression of the $h\to\gamma\gamma$ decay.} 
Let us explore now for which region of the relevant supersymmetric parameters, i.e.\  $M_2$, $\mu$ and $\tan\beta$, can $A^h_{\chi^\pm}$ have a substantial impact on the total.

\begin{figure}[t]
\centering
\includegraphics[width=0.47\textwidth,angle=0]{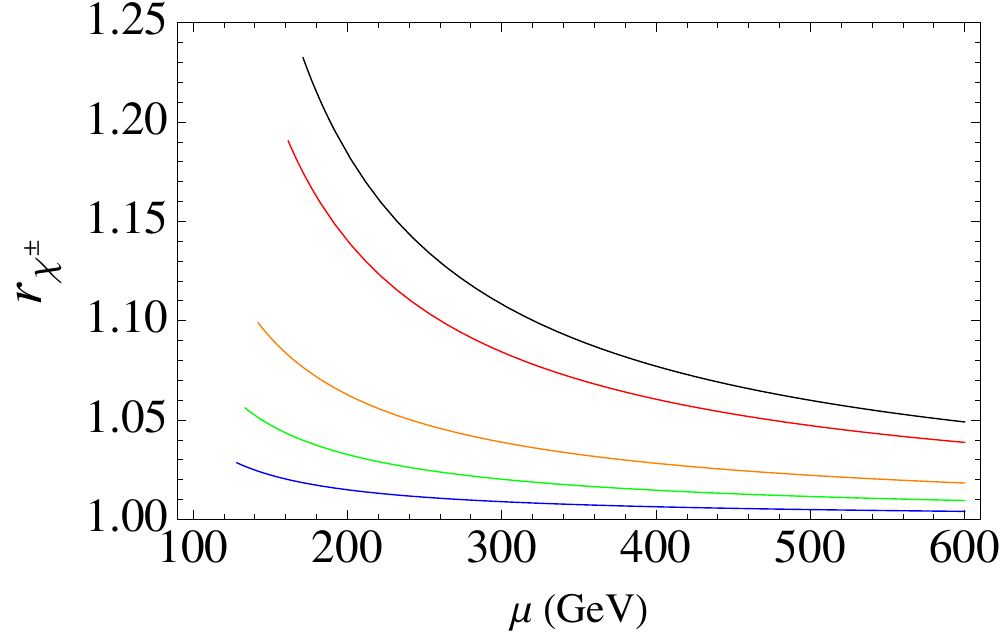}
\includegraphics[width=0.47\textwidth,angle=0]{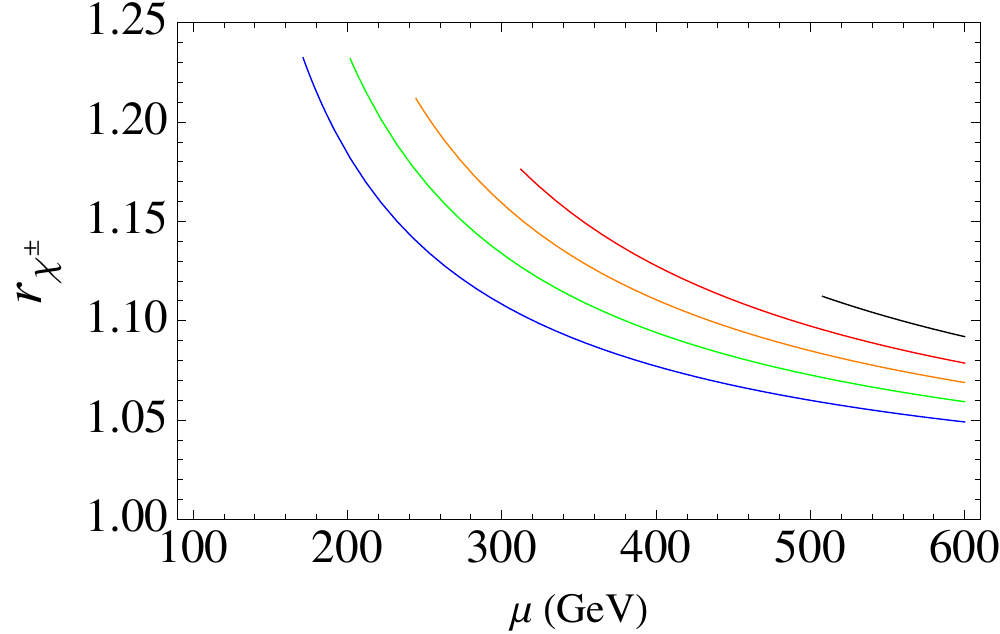}
\caption{Chargino enhancement of the $h\to\gamma\gamma$ rate, $r_{\chi^\pm}$, eq.~\eqref{eq:rchirate} vs. $\mu$. \textbf{Left}: fixing $M_2=200$ GeV while varying $\tan\beta$ to 1, 2, 5, 10 and 25, shown in black (upper line), red, orange, green, and blue (lower line). \textbf{Right}: fixing $\tan\beta=1$ while varying $M_2$ to 120, 135, 150, 170 and 200 GeV, in the same colour order. The curves are cut-off when they conflict the LEP limit $m_{\chi^\pm_1}\gtrsim 104$ GeV.}
\label{fig:Asusy}
\end{figure} 

\begin{figure}[ht]
\centering
\includegraphics[width=0.45\textwidth,angle=0]{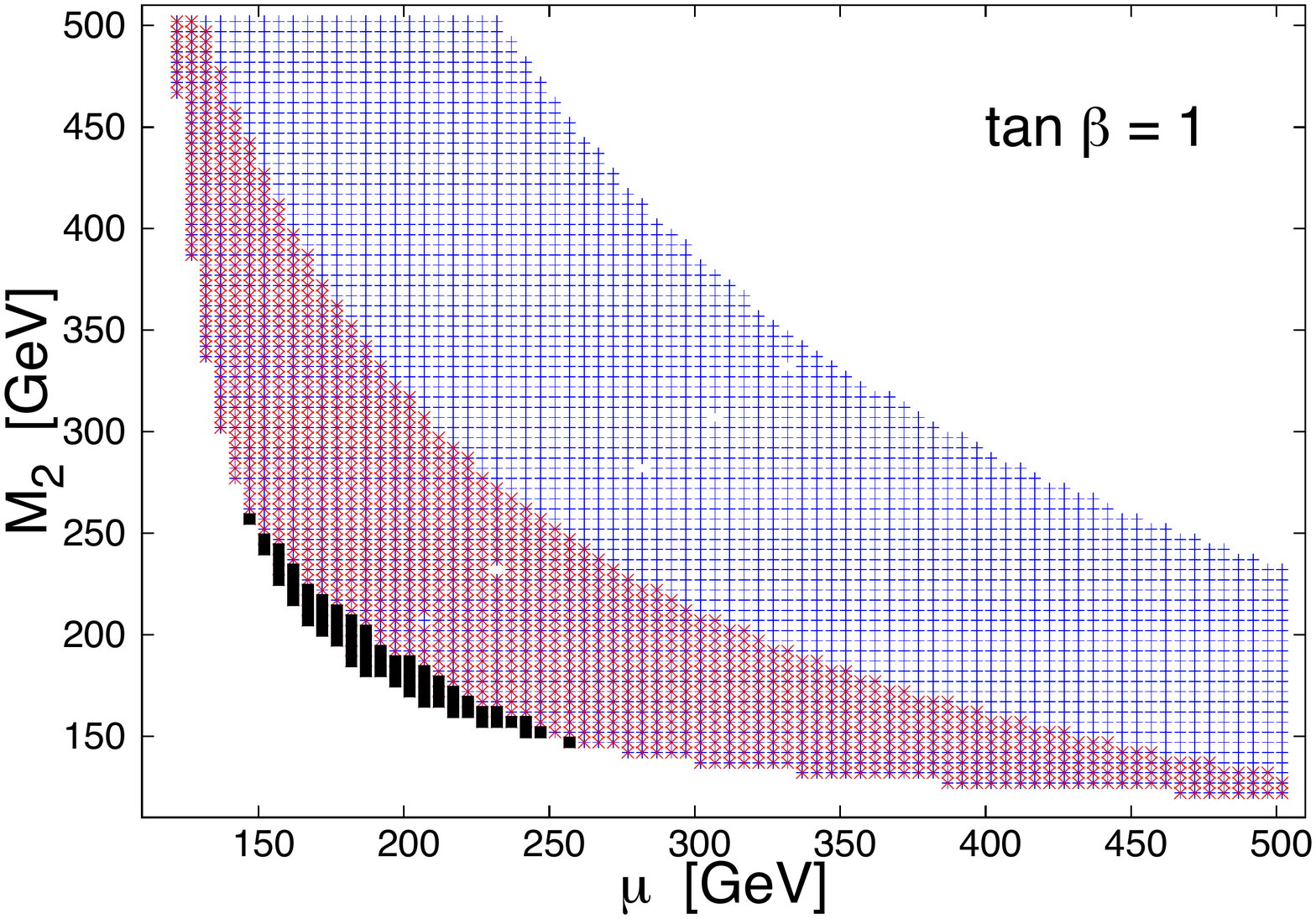}
\includegraphics[width=0.45\textwidth,angle=0]{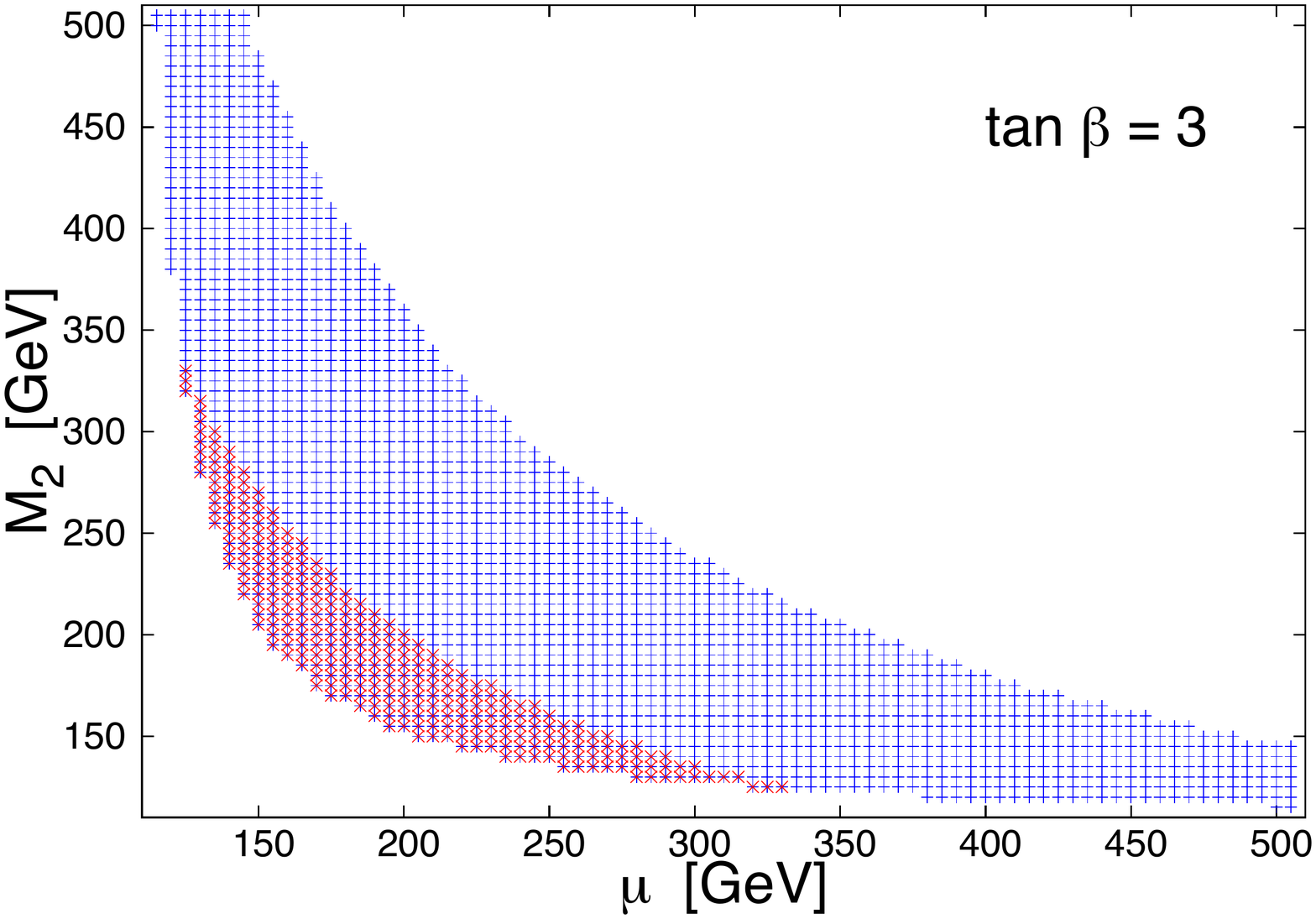}
\includegraphics[width=0.45\textwidth,angle=0]{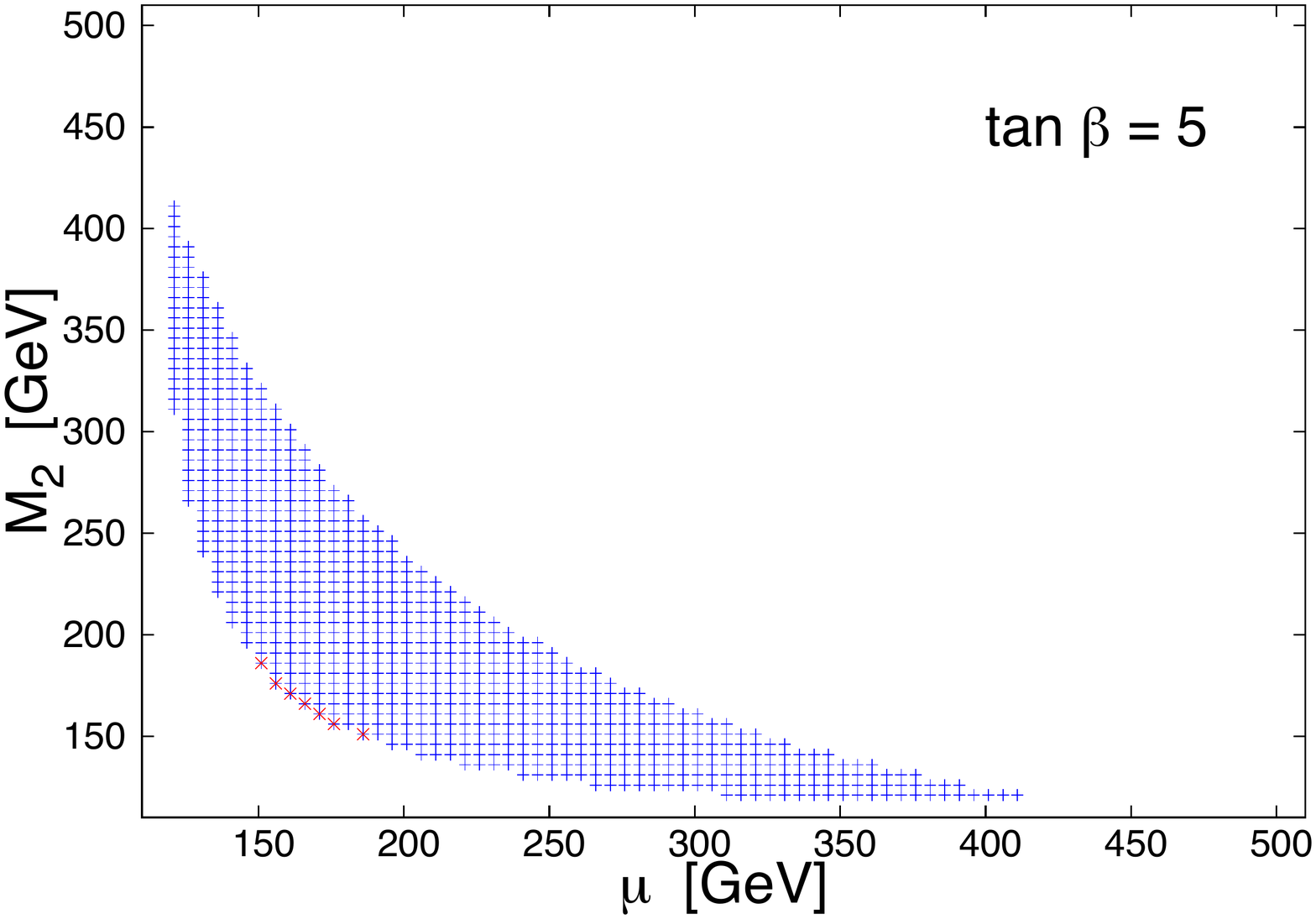}
\caption{ Regions in the $\mu$--$M_2$ plane for which $r_{\chi^\pm}\geq1.05$ (blue), $r_{\chi^\pm}\geq1.10$ (red) and $r_{\chi^\pm}\geq1.20$ (black) for  $\tan\beta=1,3,5$.}
\label{fig:triangles}
\end{figure}

Figure~\ref{fig:Asusy} shows the dependence of the enhancement rate,
\be \label{eq:rchirate}
r_{\chi^\pm}\equiv\left| \df{A^h_{SM} + A^h_{\chi^\pm} }{A^h_{\rm SM}}\right|^2 ~,
\ee
with $\mu$, for different choices of $\tan\beta$ and $M_2$. The LEP bound on the lightest chargino mass has also been included in the plots. Clearly, lower values of $\tan\beta$ and $M_2$ favour large chargino contributions, in particular the maximum is reached for $\tan\beta=1$ \cite{Huo:2012tw}. Figure~\ref{fig:triangles} shows the regions in the $\mu$--$M_2$ plane where $r_{\chi^\pm}\geq1.05, 1.1, 1.2$, i.e. more than 5\%, 10\% and 20\% enhancement, for $\tan\beta=1,3,5$. As mentioned above, the chargino contribution decreases quickly with increasing $\tan\beta$, so the $r_{\chi^\pm}\geq1.2$ region disappears for moderate values of $\tan\beta$, though the  interesting $r_{\chi^\pm}\geq1.1$ region remains. 

The point in the MSSM parameter space that presents maximum enhancement, consistent with the LEP lower bound on the chargino mass, $m_{\chi^\pm_1}\geq 104$ GeV, corresponds to $\tan\beta = 1$, $\mu = M_2 \sim 185~{\rm GeV}$, giving $r_{\chi^\pm}\simeq 1.23$.\footnote{As we will see in the next section, if the bino-mass parameter, $M_1$, is large, then there is a quasi-degeneracy between the lightest chargino and neutralino. In that case, the LEP lower bound for $m_{\chi^\pm_1}$ decreases till $\sim 95$~GeV~\cite{Beringer:1900zz}, so that the choice $\tan\beta=1$, $\mu=M_2=175$ GeV is allowed. Such model leads to a 28\% enhancement of $h\to\gamma\gamma$, which would be the absolute maximum in the MSSM.} On the other hand, the maximum $\tan\beta$ allowing for $r_{\chi^\pm}\geq1.1$ is $\tan\beta\sim5$ with $\mu$ and $M_2$ around 165 GeV.

\section{Representative models\label{sec:3}}

Here we define five representative models to illustrate the DM and LHC phenomenology of the regions of the MSSM parameter space where a relevant enhancement of the $h\to\gamma\gamma$ rate takes place (see figure~\ref{fig:triangles}).

\subsection*{Model \#1}

This model is defined close to the above-mentioned point of maximum $h\to\gamma\gamma$ enhancement, namely
\be
\tan\beta = 1,~~~ \mu = M_2 = 185~{\rm GeV}\,.
\label{model1}
\ee 
The corresponding chargino masses and  the enhancement of $h\to\gamma\gamma$ are 
\be
m_{\chi^\pm_1} = \mu-m_W = 105~{\rm GeV},~~~ m_{\chi^\pm_2} = \mu+m_W = 265~{\rm GeV}~,~~~r_{\chi^\pm}\simeq 1.23\;.
\label{mCi}
\ee
Note that the lightest chargino is close to the LEP bound (104 GeV). 

The fact that $\tan\beta$ is small demands large radiative corrections to reproduce the measured Higgs mass. Namely, one needs stop masses around $10^{6-10}$ GeV \cite{Cabrera:2011bi,Giudice:2011cg} (the magnitude depends on the size of the stop-mixing parameter, the precise value of the top mass and the size of the gluino mass). Thus the model is reminiscent of split-SUSY models, where the supersymmetric fermionic spectrum is much lighter than the scalar one. Obviously, such a scenario cannot be considered ``natural" in the sense of non-fine-tuned EW breaking. However it can be natural in a more generic sense, in particular in landscape scenarios and the like. Models of this kind have been recently considered in ref.~\cite{Ibanez:2013gf}, where $\tan\beta=1$ arises as the most natural value.

Note that the phenomenology associated to the chargino and neutralino sectors in this kind of scenarios (reviewed in section~\ref{sec:constraints}) may be the only hope to discover them at the LHC or XENON (with the possible exception of gluino detection if the gluino is also light).

\subsection*{Models \#2, \#3, \#4}

These models are representative of small $\tan\beta$ and heavier chargino masses; hence the increase in $h\to\gamma\gamma$ becomes more moderate, around 15\%. To pick up different possible patterns of gaugino-higgsino mixing (and thus chargino and neutralino couplings, and DM and LHC phenomenology), we have selected three points in parameter space:
\bea
&{\rm Model ~\#  2:}&\hspace{1.5cm}\tan\beta = 1,~~~ \mu = M_2 = 220~{\rm GeV}\,,
\nonumber\\
&{\rm Model ~\#  3:}&\hspace{1.5cm}\tan\beta = 1,~~~ \mu = 200~{\rm GeV},\ M_2 = 250~{\rm GeV}\,,
\nonumber\\
&{\rm Model ~\#  4:}&\hspace{1.5cm}\tan\beta = 1,~~~ \mu = 250~{\rm GeV},\ M_2 = 200~{\rm GeV}\,.
\nonumber
\eea
The corresponding chargino spectrum and  $h\to\gamma\gamma$ enhancement are
\bea
&{\rm Model ~\#  2:}&\hspace{1.5cm}m_{\chi^\pm_1} = 139~{\rm GeV},~~~ m_{\chi^\pm_2} = 301~{\rm GeV},~~~ r_{\chi^\pm}\simeq 1.14\,,
\nonumber\\
&{\rm Model ~\#  3:}&\hspace{1.5cm}m_{\chi^\pm_1} = 141~{\rm GeV},~~~ m_{\chi^\pm_2} = 309~{\rm GeV},~~~ r_{\chi^\pm}\simeq 1.14\,,
\nonumber\\
&{\rm Model \#  4:}&\hspace{1.5cm}m_{\chi^\pm_1} = 141~{\rm GeV},~~~ m_{\chi^\pm_2} = 309~{\rm GeV},~~~ r_{\chi^\pm}\simeq 1.14\,.
\nonumber
\eea
Note that the models have been chosen so that they present similar chargino masses and $h\to\gamma\gamma$ enhancement.
The remaining characteristics are as for Model \#1.

\subsection*{Model \#5}

This model is defined to be close to the maximum value of $\tan\beta$ one can afford while keeping a substantial $h\to\gamma\gamma$ enhancement (around 10\%). The parameters are:
\be
\tan\beta = 5,~~~ \mu = M_2 = 165~{\rm GeV}\,.
\ee 
The corresponding chargino masses and $h\to\gamma\gamma$ enhancement are:
\be
m_{\chi^\pm_1} = 104~{\rm GeV},~~~ m_{\chi^\pm_2} = 238~{\rm GeV},~~~ r_{\chi^\pm}\simeq 1.10\,.
\ee
The fact that $\tan\beta$ is moderate allows for stop masses of the usual size $\mathcal O$(TeV) in order to reproduce the Higgs mass. Hence, taking into account the existing LHC limits, the whole SUSY-QCD sector can be as light as order 1--2~TeV.

\section{Dark Matter bounds\label{sec:4}}

The neutralino mass matrix is given by
\begin{displaymath}
M_N = \left(
\begin{array}{cccc}
M_1 &  0 & - m_Z s_W c_\beta & m_Z s_W s_\beta \\
0 & M_2  & m_Z c_W c_\beta & - m_Z c_W s_\beta \\
 - m_Z s_W c_\beta & m_Z c_W c_\beta & 0 & -\mu \\
 m_Z s_W s_\beta & - m_Z c_W s_\beta & -\mu & 0 
\end{array}
\right)\,,
\label{MNmat}
\end{displaymath}
so the neutralino sector of the theory is determined by the same parameters that define the chargino sector ($\tan\beta$, $M_2$, and $\mu$) plus the bino mass parameter, $M_1$.  Usually, the lightest neutralino, $\chi^0_1$, is the lightest supersymmetric particle (LSP) and a good candidate to account (at least partially) for the dark matter of the universe.
In consequence DM searches are likely to put important bounds on the scenarios we have described in the previous sections.

There are two main DM constraints to consider. First, assuming that the LSP has been produced thermally in the early universe, one has to check that its present abundance coincides with the DM one, $\Omega h^2 \approx 0.11$, 
or at least is lower than that (in the latter case there should be other DM species, e.g.\ axions). Second, direct detection experiments, in particular XENON100, put limits on the DM-nucleon elastic cross section, given that such processes have not been bp observed so far.\footnote{ See however ref.~\cite{Agnese:2013rvf}.} At present, the strongest bound on the spin-independent cross-section is $2\times10^{-9}$~pb for a DM mass of about 55~GeV.

In addition, there are further constraints on the neutralino sector. In particular, the $h\to\chi^0_1\chi^0_1$ process contributes to the invisible width of the Higgs, which is at present bounded from above at roughly 20\%~\cite{Falkowski:2013dza,Giardino:2013bma,Belanger:2013kya,Djouadi:2013qya}.

Next, we analyse all these constraints in detail for Model \#1. For the other representative models the analysis and results present very similar features, so we discuss them afterwards in a briefer fashion.

\vspace{0.2cm}
\noindent 
{\underline {\em Model \#1}}

\vspace{0.2cm}

\noindent
The fact that in this model $\mu=M_2$ and $\tan\beta=1$, see eq.~(\ref{model1}), allows for the analytic diagonalization of the neutralino mass matrix in  two useful limits of $M_1$, which is the only relevant free parameter:

\vspace{0.2cm}
\underline{$|M_1|\ll\mu=M_2$}
\bea
&&
m_{\chi^0_1} = M_1 - \df{m_Z^2 s^2_W}{\mu} +{\cal O}\left( \df{m_Z^2 s^2_W M_1}{\mu^2}\right),~~
m_{\chi^0_2} =\mu-m_W + \df{m_Z^2 s^2_W}{2\mu} +{\cal O}\left( \df{m_Z^2 s^2_W M_1}{\mu^2}\right),~~ \nonumber \\
&&
m_{\chi^0_3} =\mu\,,~~
m_{\chi^0_4} = \mu + m_W + \df{m_Z^2 s^2_W}{2\mu} +{\cal O}\left( \df{m_Z^2 s^2_W M_1}{\mu^2}\right),
\label{limit1}
\eea

\begin{figure}[t]
\centering
{\includegraphics[width=0.49\textwidth,angle=0]{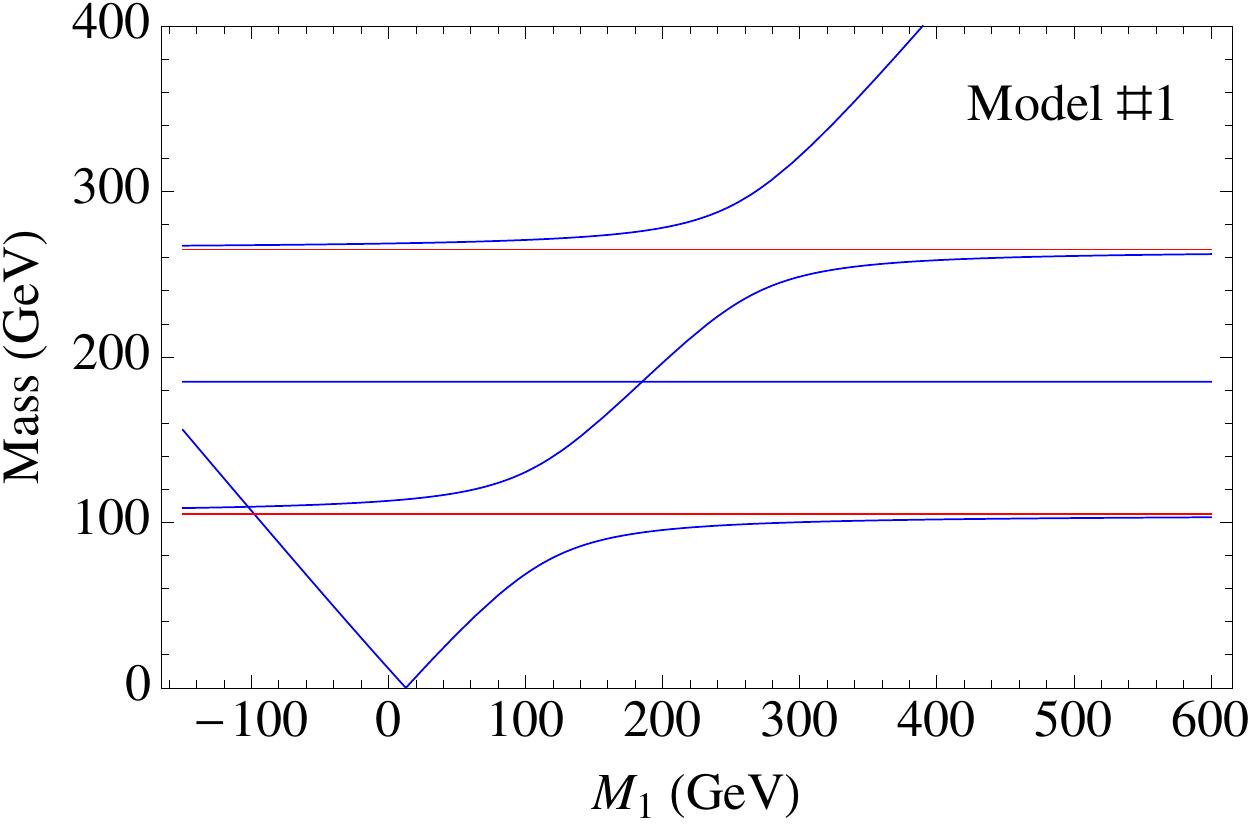}}
{\includegraphics[width=0.49\textwidth,angle=0]{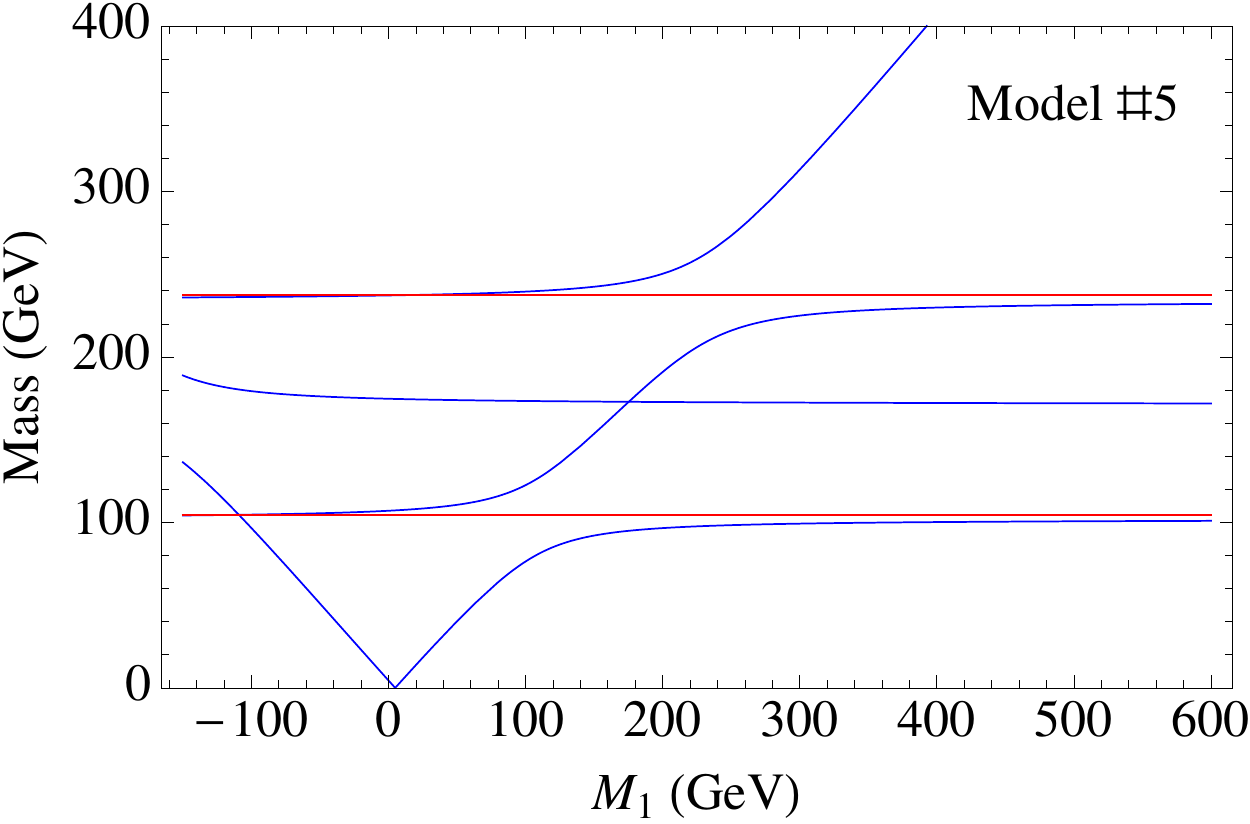}}
\caption{Masses of neutralinos (blue) and charginos (red) as functions of $M_1$ for Models \#1 (left panel) and \#5 (right panel).
}
\label{fig:masses}
\end{figure}

\underline{$|M_1|\gg \mu=M_2$}
\bea
\label{limit2}
&&
m_{\chi^0_1} = \mu - m_W - \df{m_Z^2 s^2_W}{2M_1} +{\cal O}\left( \df{m_Z^2 s^2_W \mu}{M_1^2}\right),~~
m_{\chi^0_2} =\mu\,, \\
&&
m_{\chi^0_3} =\mu + m_W  - \df{m_Z^2 s^2_W}{2M_1} +{\cal O}\left( \df{m_Z^2 s^2_W\mu}{M_1^2}\right),~~
m_{\chi^0_4} = M_1  + \df{m_Z^2 s^2_W}{M_1} +{\cal O}\left( \df{m_Z^2 s^2_W\mu}{M_1^2}\right). \nonumber
\eea
Although strictly speaking eqs.~(\ref{limit1}) and~(\ref{limit2}) are only valid for $\mu=M_2$ and $\tan\beta=1$, they also give a good hint of the neutralino spectrum in the limit of small and large $M_1$ in more general cases. This is illustrated by figure ~\ref{fig:masses}, which shows the spectrum of charginos and neutralinos as a function of $M_1$ for Model \#1 and Model \#5, which are our two extreme cases of $\tan\beta$. They exhibit similar features. Namely for large $M_1$, the lightest chargino, $\chi_1^\pm$, becomes quasi-degenerate with the lightest neutralino, $\chi^0_1$,  while for small $|M_1|$ its mass gets close to that of the second lightest neutralino, $\chi^0_2$. Likewise, for $M_1 \lesssim -100$~GeV
the LSP is not $\chi^0_1$ anymore, since $m_{\chi^0_1} > m_{\chi^\pm_1}$, cf.\ eq.~\eqref{mCi}. Thus, in the following, we will scan $M_1$ in the $-100$ GeV $\lesssim M_1 \lesssim 800$ GeV range. 

Since sfermions are heavy in this scenario, they do not contribute significantly to annihilation and direct-detection cross sections. These processes occur via exchange of Higgs or $Z$ bosons, while the annihilation rate can be additionally enhanced by coannihilation with the light chargino. Hence, the relevant couplings are~\cite{Djouadi:2005gj}:
\bea
g_{h\chi^0_i\chi^0_j} &=& g^L_{h\chi^0_i\chi^0_j} P_L + g^R_{h\chi^0_i\chi^0_j} P_R~\nonumber \\
&=& \df{1}{2s_W} (Z_{j2} - \tan\theta_W Z_{j1}) (-\sin\alpha Z_{i3} -\cos\alpha Z_{i4}) + i\leftrightarrow j ~. \\ 
g_{Z\chi^0_i\chi^0_j} &=& g^L_{Z\chi^0_i\chi^0_j} P_L + g^R_{Z\chi^0_i\chi^0_j} P_R~
\nonumber\\
& =& \df{1}{2 s_W c_W} [(Z_{i4} Z_{j4} - Z_{i3} Z_{j3}) P_L - (Z_{i4} Z_{j4} - Z_{i3} Z_{j3}) P_R ]\, ,
\label{Zcouplings}
\eea
where $Z_{ij}$ is the real-valued matrix that diagonalizes the neutralino mass-matrix, $ZM_NZ^T={\rm diag}(m_{\chi^0_1},m_{\chi^0_2},m_{\chi^0_3},m_{\chi^0_4})$. It is easy to check that for $\tan \beta=1$ and $\mu=M_2$ one has $Z_{13}=-Z_{14}$, so that the coupling of $\chi^0_1$ to the $Z$ boson vanishes (except for tiny radiative corrections). Hence for Model \#1, only the coupling of the $\chi^0_1$ to the Higgs is relevant. This occurs also exactly for Model \#2 and approximately for Models \#3 and \#4. For Model \#5, since $\tan\beta=5$, the coupling to the $Z$ boson becomes relevant.

\begin{figure}[t]
\centering
{\includegraphics[width=0.49\textwidth,angle=0]{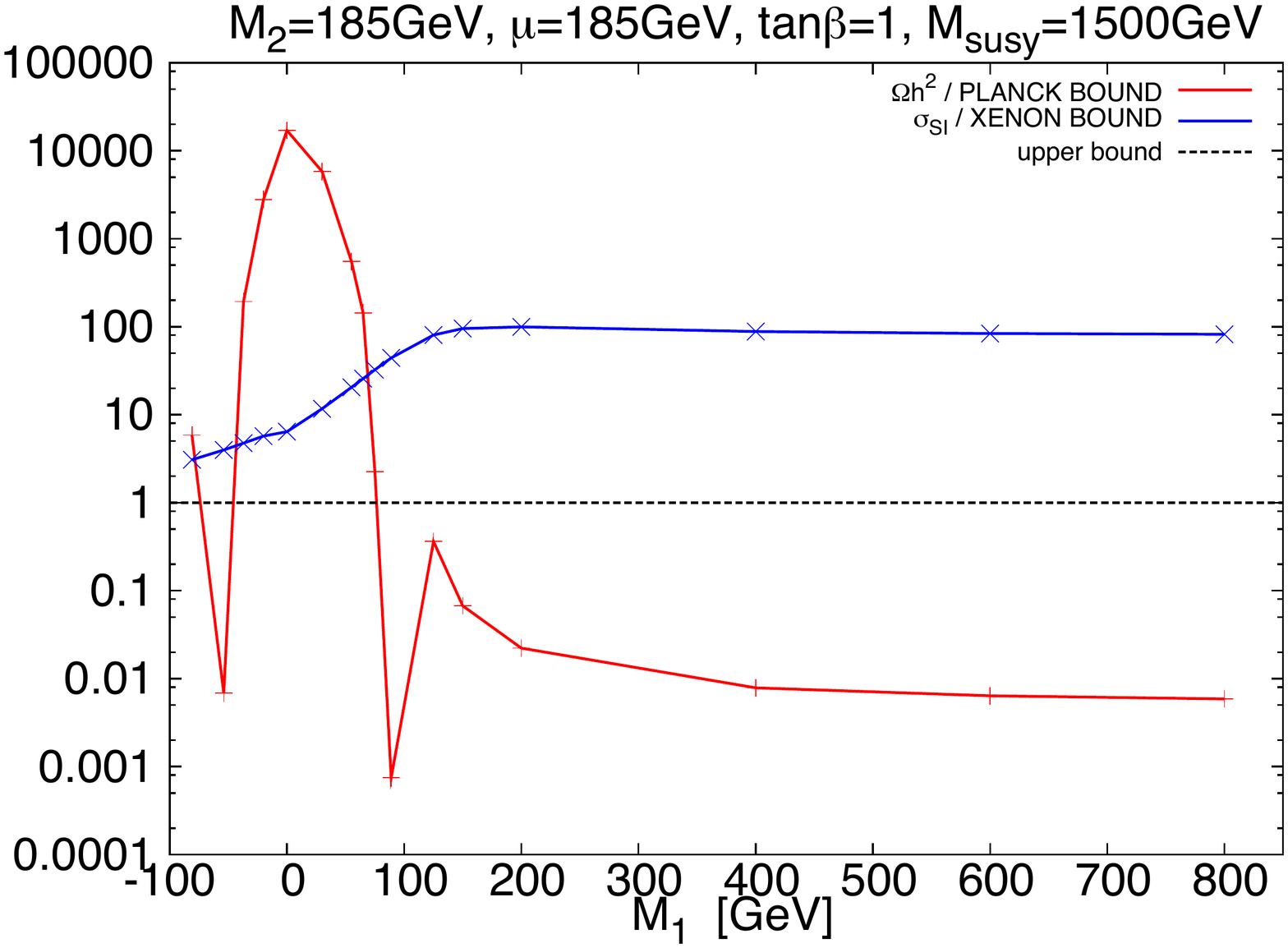}}
{\includegraphics[width=0.49\textwidth,angle=0]{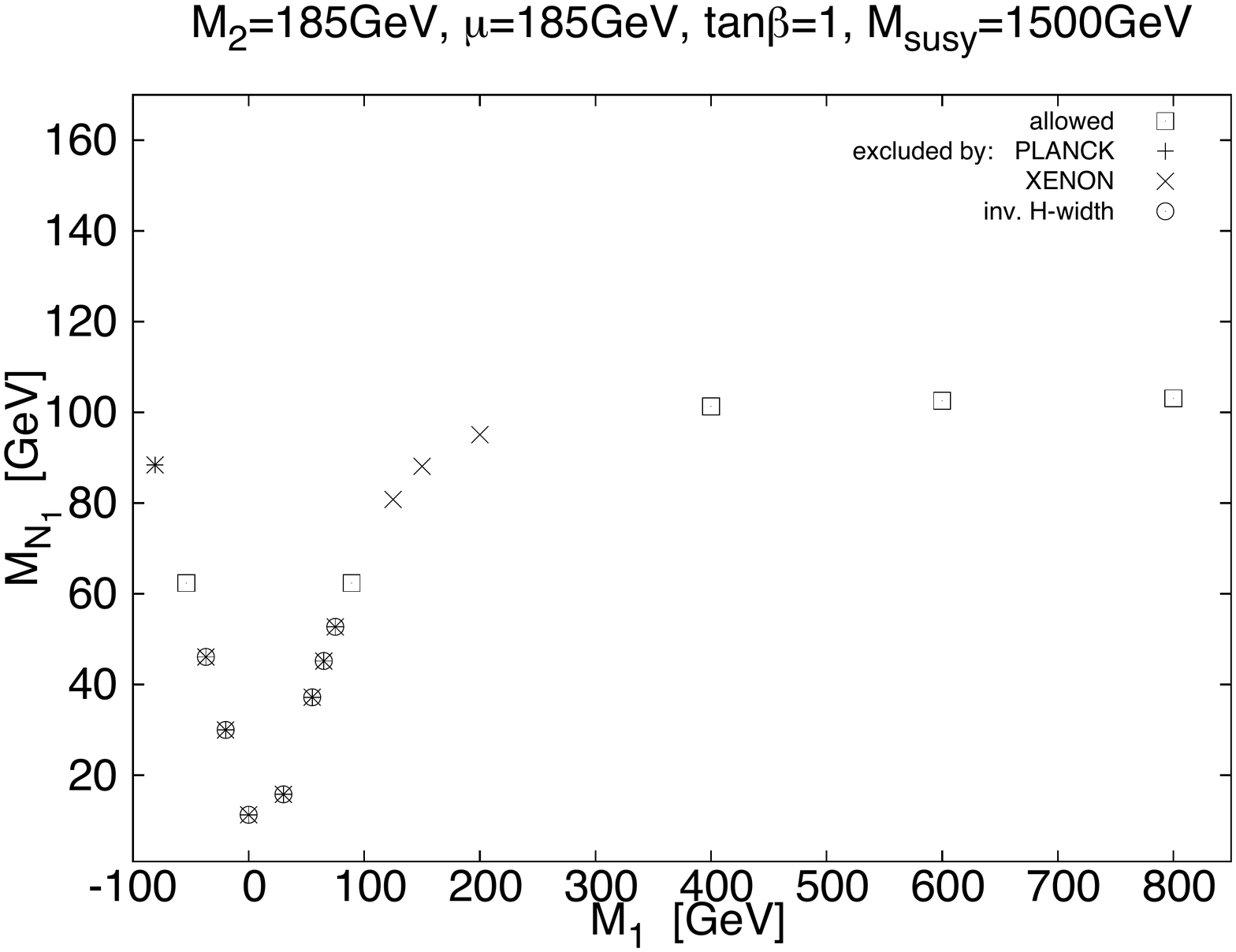}}
\caption{{\bf Left panel}: WMAP and XENON bounds  as a function of $M_1$, for Model \#1; red line: DM abundance normalised to the Planck result; blue line: spin-independent direct-detection cross section $\sigma_{\chi-n}$ over the experimental limit, $2\times10^{-9}$~pb, before normalising by the neutralino fraction of DM. {\bf Right panel}: LSP mass as a function of $M_1$. At each value of $M_1$ it is shown whether the model is allowed or excluded by experimental constraints. }
\label{fig:WX}
\end{figure}

We describe now the results corresponding to different regions of the free parameter, $M_1$. They have been obtained using the {\tt micrOmegas} code~\cite{Belanger:2010gh}, and are summarised in figure~\ref{fig:WX}. 

\begin{itemize}

\item If $\chi_1^0$ were the unique DM component in the local halo, the LSP-nucleon cross-section, $\sigma_{\chi-n}$, would be always above the experimental upper bound given by XENON (see figure~\ref{fig:WX}, blue line). This is because the coupling of $\chi_1^0$ to the Higgs is always sizeable. This is a quite general fact for all the models with very low $\tan\beta$, not just for Model \#1 (but can be avoided for larger values of $\tan\beta$, see below). The scenario can only be reconciled with DM observations when the relic $\chi^0_1$ density is sufficiently smaller than the DM one (so the neutralino cannot be the only DM component). This is actually the case in some ranges of $M_1$, as we discuss next.

\item $M_1\gg\mu,M_2$. As can be seen from the left panel of figure~\ref{fig:WX} (red line), this region leads to a suppressed contribution of the neutralinos to $\Omega h^2$, essentially because $\chi_1^0$  is very close in mass to $\chi^\pm_1$ (see figure~\ref{fig:WX}, right panel), and the corresponding co-annihilation process is very efficient to decrease its relic abundance. In consequence, the normalised $\sigma_{\chi-n}$ survives the constraints of direct detection. Besides, since in this region $m_{\chi^0_1}>m_h/2$, there is no contribution to the invisible width of the Higgs. This region of the parameter space is quite broad and can be tested by the future XENON1T, since it is not far from the experimental limit of XENON100.

\item $M_1 \approx \mu, M_2$. As shown in figure~\ref{fig:WX}, in this region the masses of the lightest neutralino and the light chargino become too split to allow for an efficient co-annihilation. Although the annihilation processes (mediated by the Higgs) are still efficient to decrease the neutralino abundance below the DM one, the corresponding direct-detection rate is too large and the region becomes excluded by the direct-detection bound, even after normalising it to the  DM fraction of neutralinos.

\item $|M_1|\lesssim 100$ GeV. In this region there are two windows (for positive and negative $M_1$) for which $m_{\chi^0_1}\approx m_h/2$, and the LSP annihilation (through the Higgs in the $s$-channel) is close to the resonance. This causes the relic density to decrease abruptly, thus normalising the direct detection bound to allowed values (see the corresponding points in figure~\ref{fig:WX}).    

\item $|M_1| \ll \mu, M_2$. As shown in figure~\ref{fig:WX}, this region leads to excessive relic density. This happens, essentially, because the LSP is mostly bino and thus its annihilation is very inefficient. In addition, since $m_{\chi^0_1}<m_h/2$ the Higgs decay to LSPs is possible and, actually, it exceeds the bound on invisible width of the Higgs~\cite{Falkowski:2013dza,Giardino:2013bma,Belanger:2013kya,Djouadi:2013qya}.

\end{itemize}

\vspace{0.2cm}
\noindent 
{\underline {\it Models \#2, \#3, \#4, \#5}}

\vspace{0.2cm}

\begin{figure}[hbt]
\centering
\includegraphics[width=0.49\textwidth,angle=0]{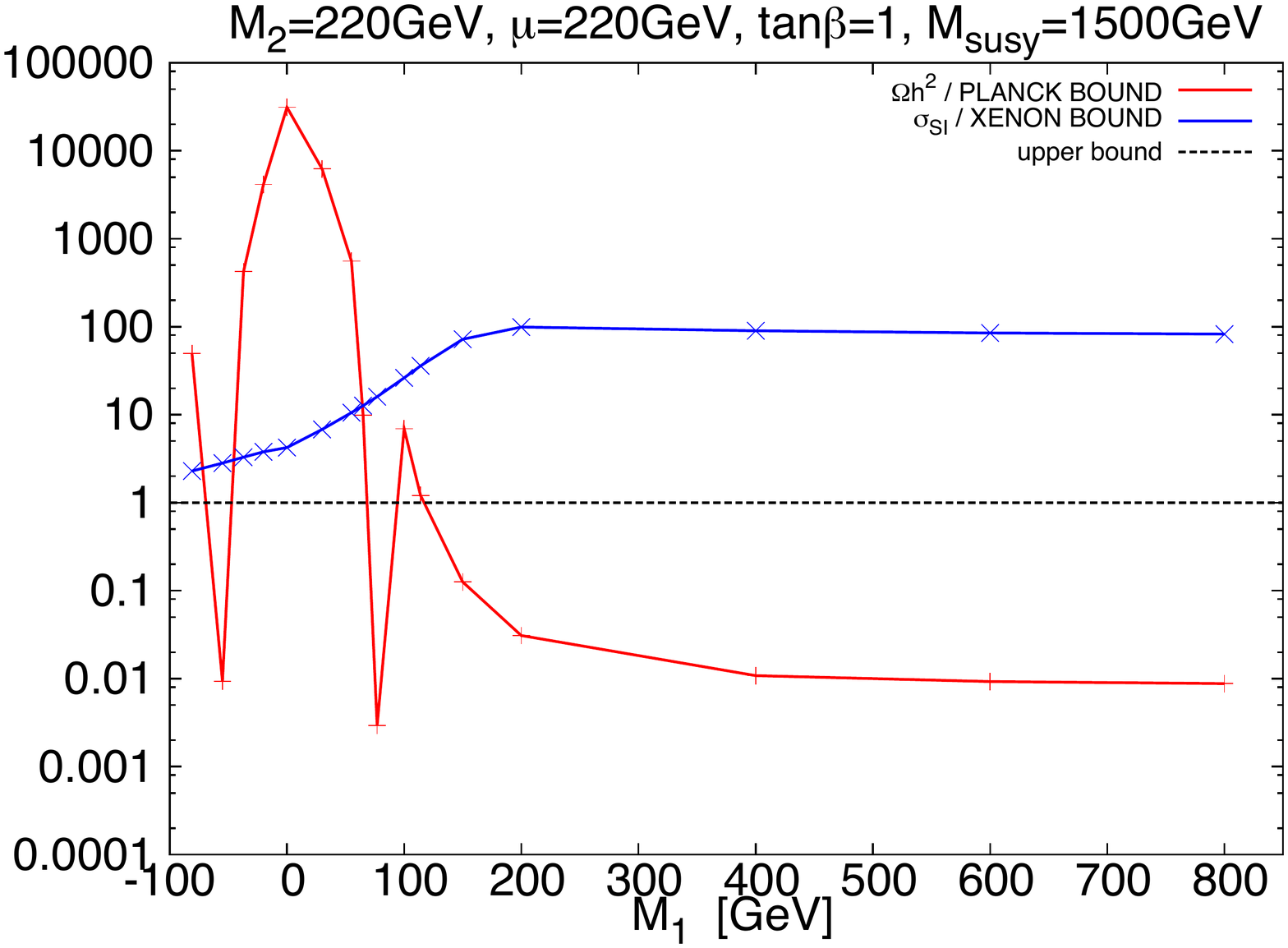}
\includegraphics[width=0.49\textwidth,angle=0]{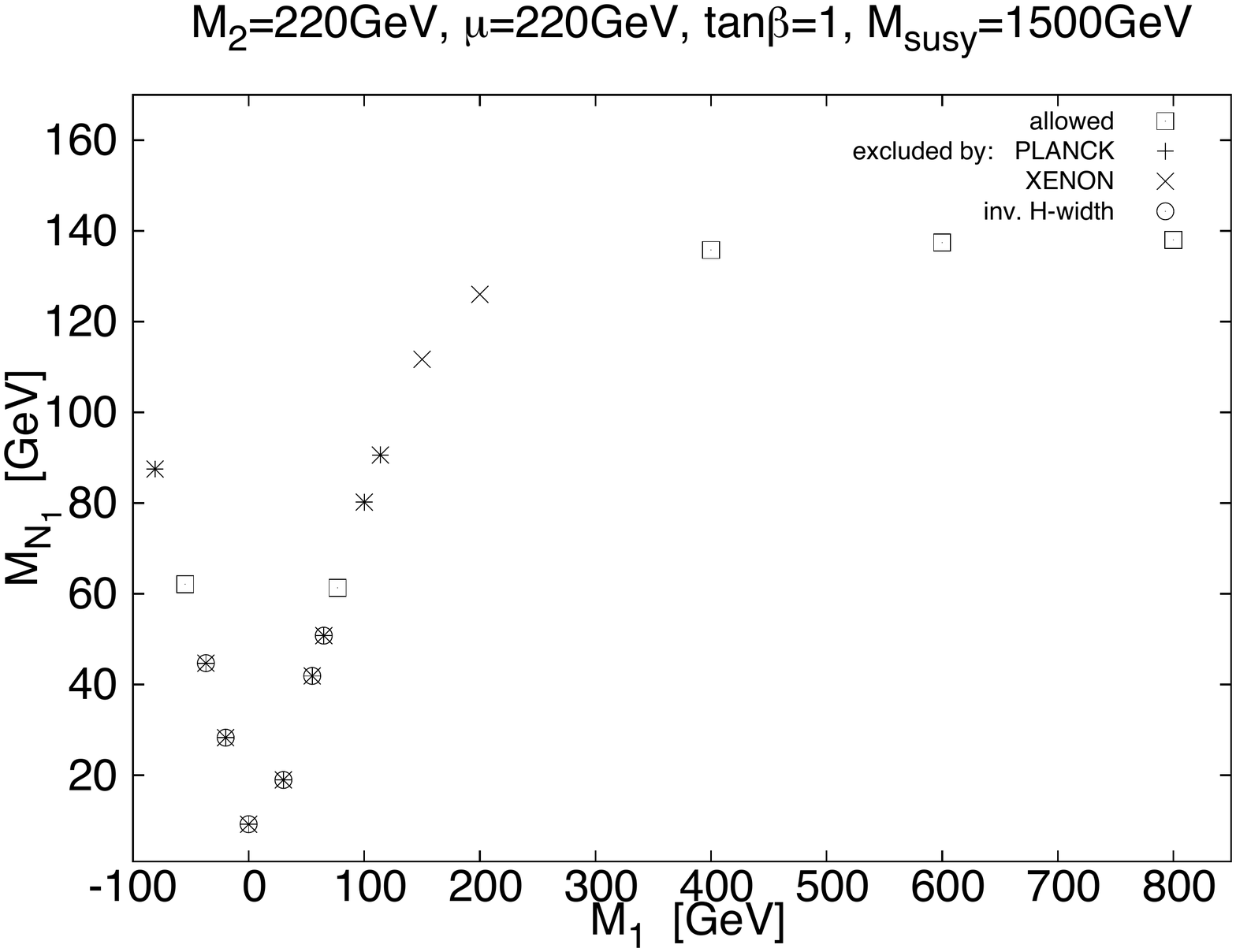}
\includegraphics[width=0.49\textwidth,angle=0]{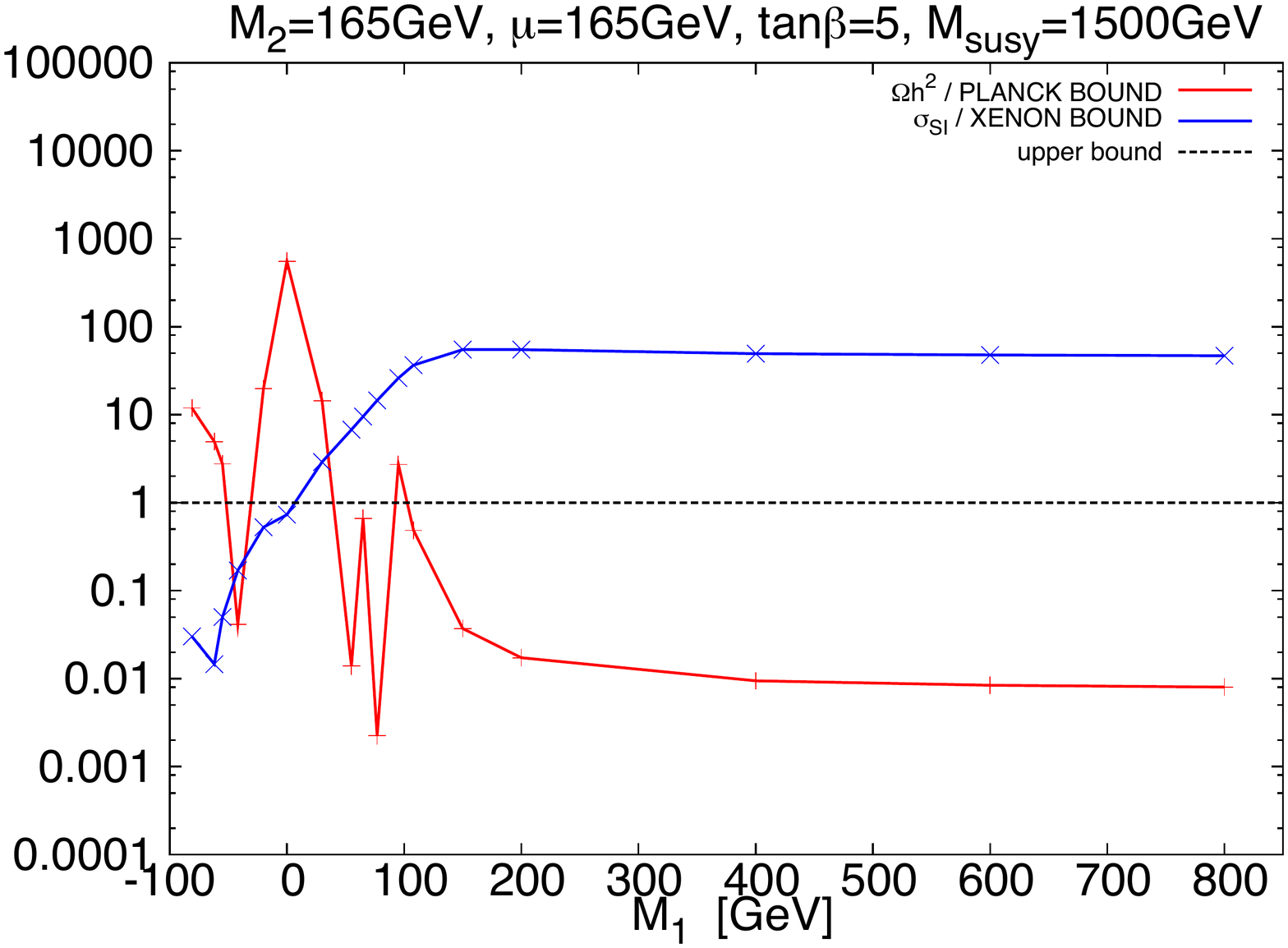}
\includegraphics[width=0.49\textwidth,angle=0]{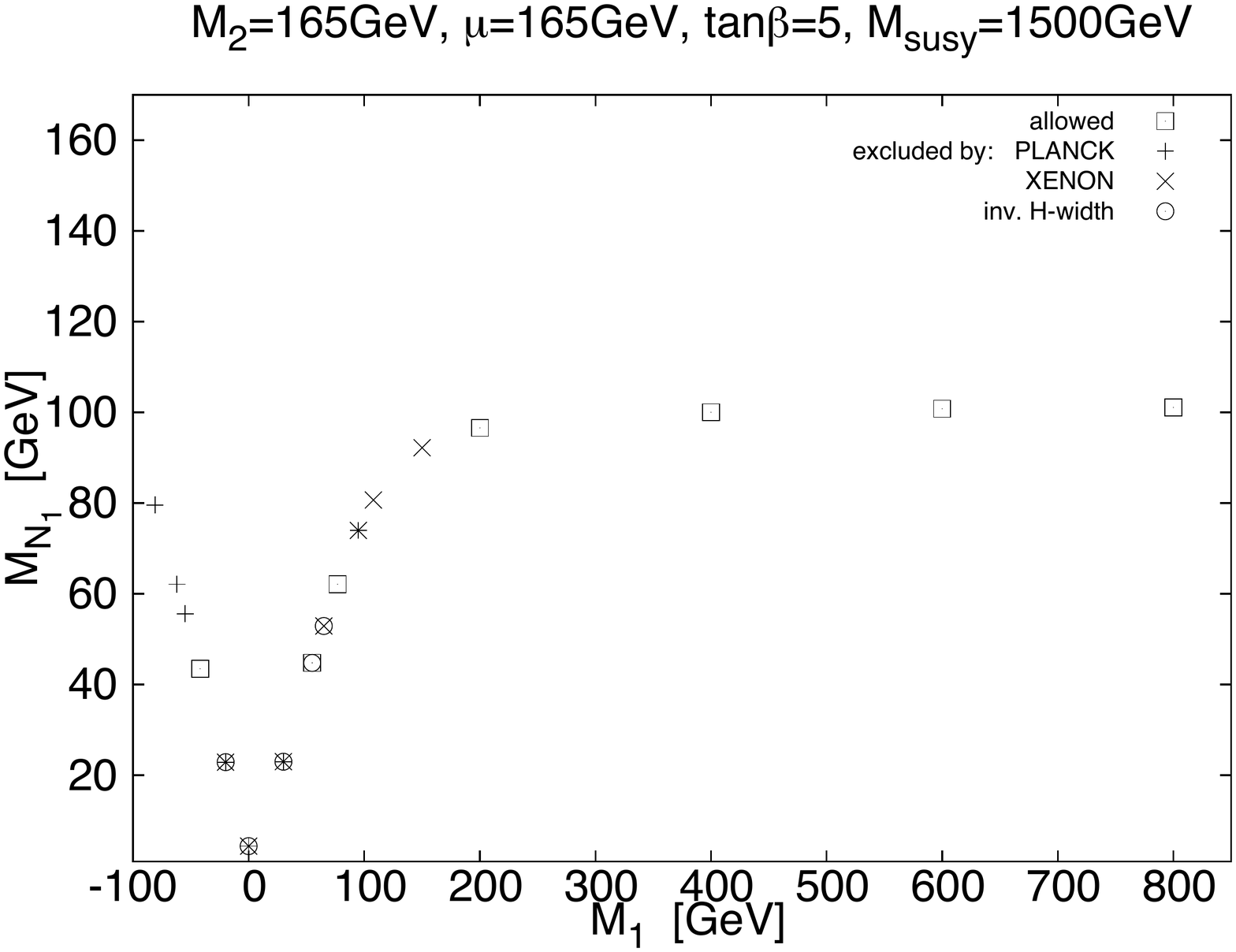}
\caption{ \label{fig:WXrest}  As figure \ref{fig:WX} but for Models \#2, \#3, \#4 (upper panels) and Model \#5 (lower panels). Models \#2, \#3, \#4 have a very similar performance and have been presented in a unified way.
 }
\end{figure}

\noindent
As commented above, these models present similar DM features as Model \#1. The results are summarised in figure~\ref{fig:WXrest}, which are analogous to the previous figure~\ref{fig:WX}. The only remarkable differences occur for Model \#5. Since  the neutralino coupling to the $Z$ boson, eq.~(\ref{Zcouplings}), increase with $\tan\beta$, for Model \#5 (which has $\tan\beta=5$) it competes with the Higgs-neutralino coupling. As a consequence, for $|M_1|\lesssim 100$ GeV there appears a window at $m_{\chi^0_1}\approx m_Z/2$, where the LSP annihilation occurs through a $Z$ in the $s$-channel, which is however excluded by the invisible width of the Higgs. In addition, for negative $M_1$, the direct detection (spin-independent) cross section drops drastically, well below the experimental limit. This is due to an accidental cancellation in the Higgsino contribution to the neutralino-Higgs coupling.\footnote{The $Z$ exchange in the neutralino-nucleon interaction does not contribute to the spin-independent cross-section.} As a consequence, the possibility of getting the observed DM abundance becomes now feasible, as it is clear from figure~\ref{fig:WXrest} (this happens for $M_1\simeq -49.5$ GeV).

\section{Electroweak observables\label{sec:STU}}

Since we are dealing with  charginos and neutralinos close to the electroweak scale, we should consider their impact on the electroweak precision data~\cite{Martin:2004id}.  A good estimate of this can be done by evaluating the corrections to the oblique parameters $S$, $T$ and $U$  \cite{Peskin:1991sw}, whose present ($1\sigma$) experimental values are \cite{Baak:2012kk}:
\bea
S  &=&  0.03 \pm 0.10~, \nonumber \\
T  &=&  0.05 \pm 0.12~, \nonumber\\ 
U  &=&  0.03 \pm 0.10~.
\label{STU}
\eea
The correlation coefficients can also be found in ref.~\cite{Baak:2012kk}. These results are similar to those of \cite{Erler:2013xha}.

\begin{figure}[t!]
\centering
\includegraphics[width=0.8\textwidth,angle=0]{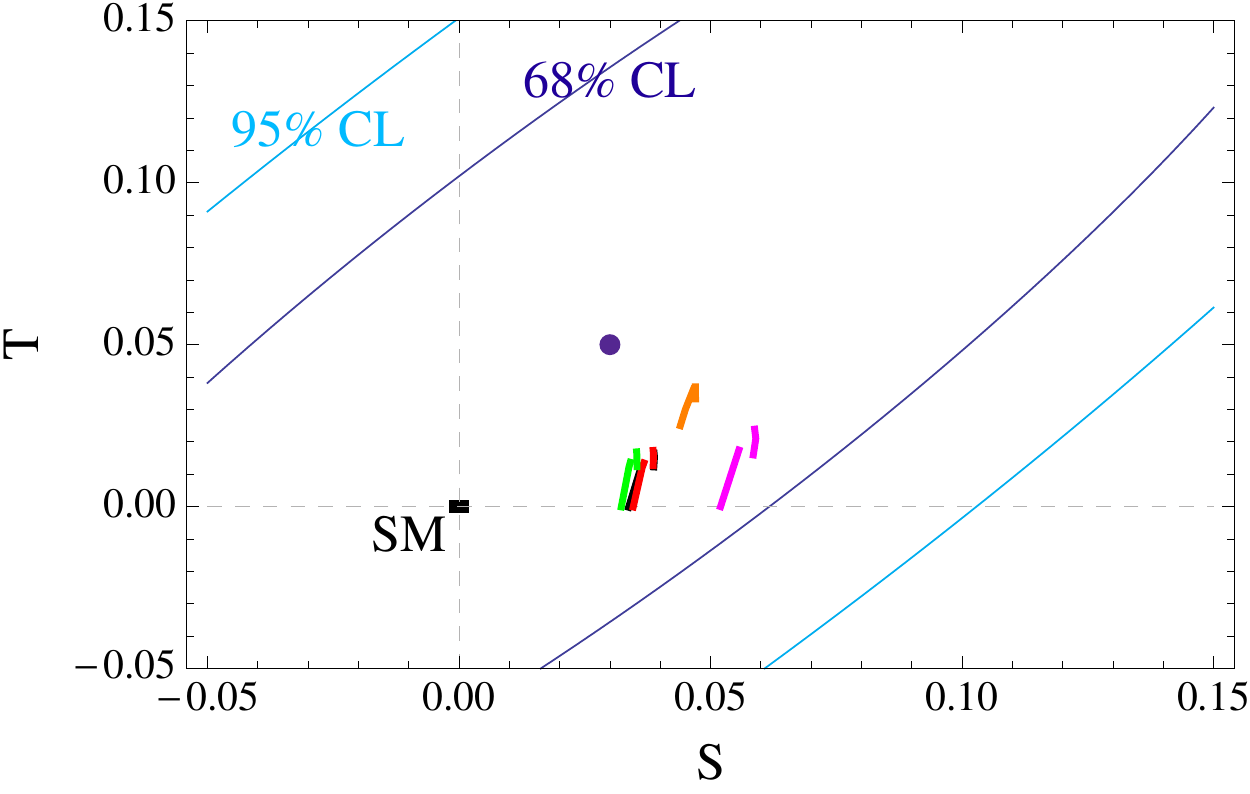} 
\caption{$S$ and $T$ parameters for benchmark Models \#1(magenta), \#2 (red), \#3 (black), \#4 (green) and \#5 (orange) when $M_1$ is varied in the range $-100$ GeV $\lesssim M_1 \lesssim 800$ GeV, requiring $m_{\chi_1^0}\geq M_Z/2$. Models \#2, \#3 and \#4 appear greatly overlapped. For each model there are an upper and a lower subsegment, corresponding to the negative and positive ranges of $M_1$. The SM prediction is the black rectangle at $(0,0)$. The experimental results are given by the 68\% C.L. (deep blue) and 95\% C.L. (light blue) elliptical lines around the central values at the blue dot.}
\label{fig:STU}
\end{figure}

We have computed the supersymmetric corrections to $S$, $T$ and $U$ for the five benchmark models, finding that they are in all instances inside the $1\sigma$ ranges. In some cases, the supersymmetric corrected values get even closer to the experimental values than the SM result. This is shown in figure~\ref{fig:STU}, where, following ref. \cite{Baak:2012kk}, we have plotted the 68\% and 95\% CL regions in the $S$, $T$ plane leaving $U$ unconstrained. The segments are spanned by the change in $S$ and $T$ for each benchmark model when $M_1$ is varied within the scanned range, $-100$ GeV $\lesssim M_1 \lesssim 800$ GeV, requiring $m_{\chi_1^0}\geq M_Z/2$. The upper and lower subsegments for each model correspond to the negative and positive $M_1$ subranges respectively, and their upper ends are always located at $m_{\chi_1^0}= M_Z/2$. Notice that the SM point lies at $(S,T) = (0,0)$ since the SM contribution (evaluated at $m_t= 173$ GeV, $m_h=126$ GeV) is substracted in the definition of the oblique parameters.

\section{Collider constraints and detection prospects\label{sec:constraints}}

\begin{figure}[t!]
\centering
\includegraphics[width=0.4\textwidth,angle=0]{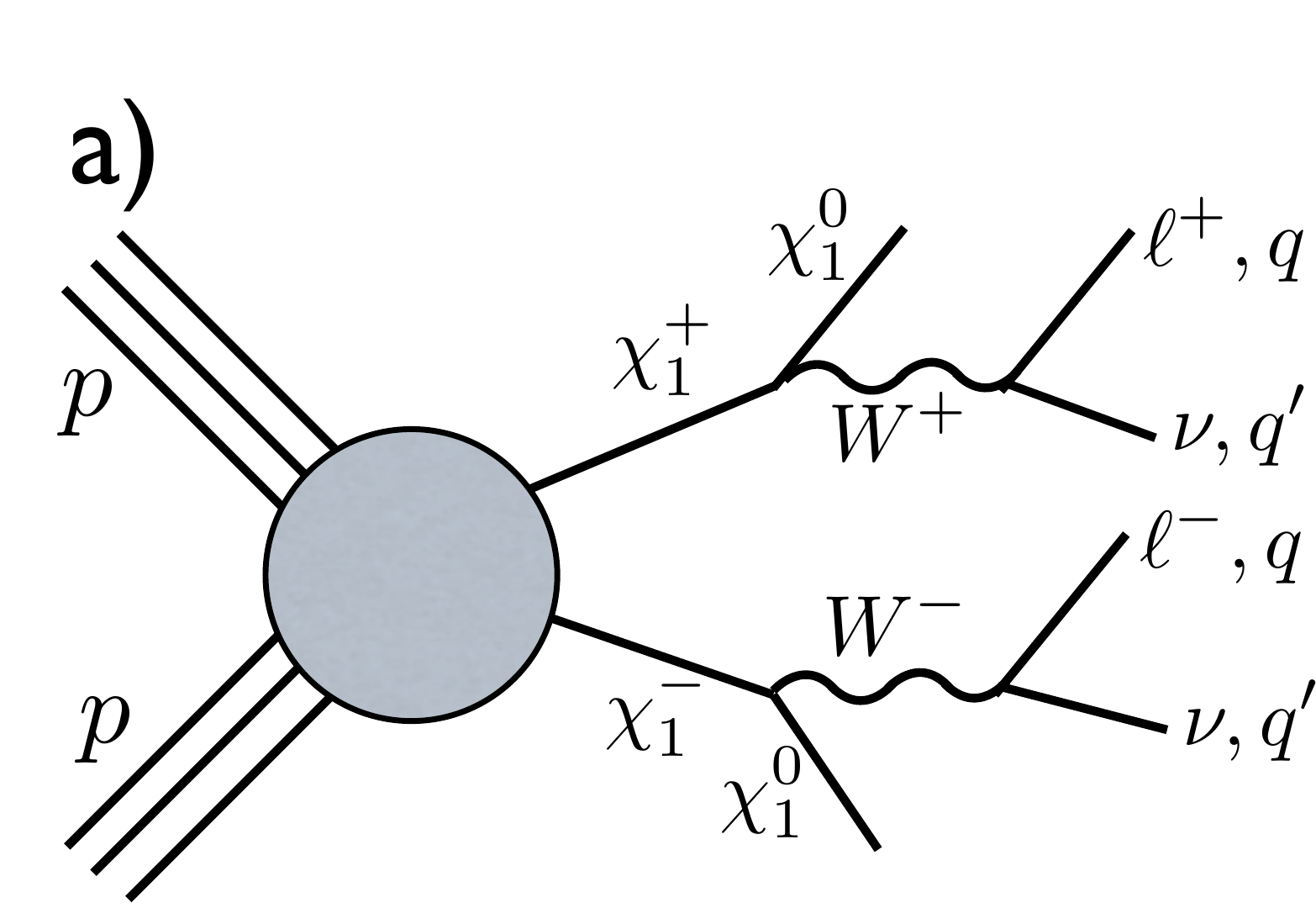} 
\includegraphics[width=0.4\textwidth,angle=0]{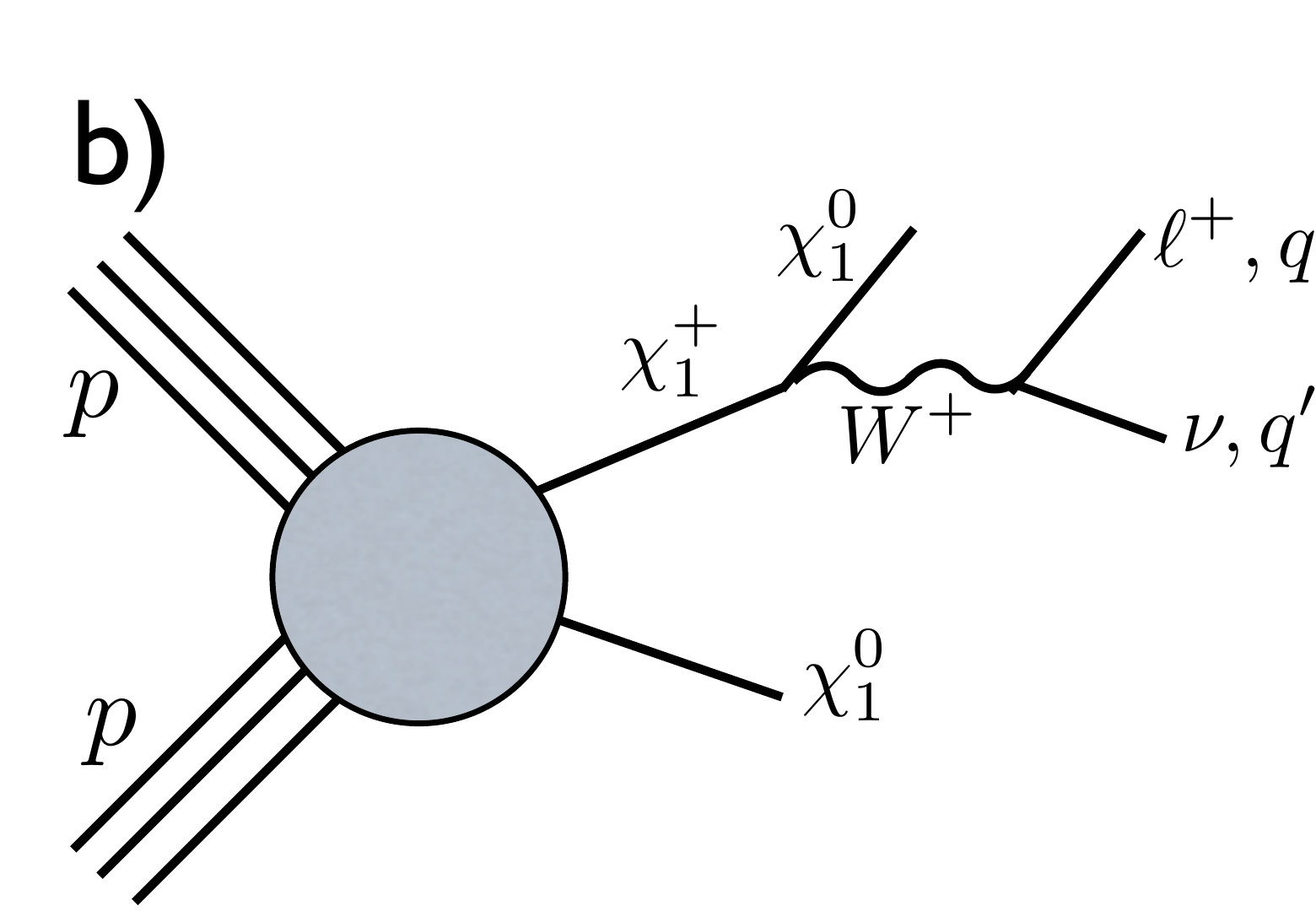} 
\includegraphics[width=0.4\textwidth,angle=0]{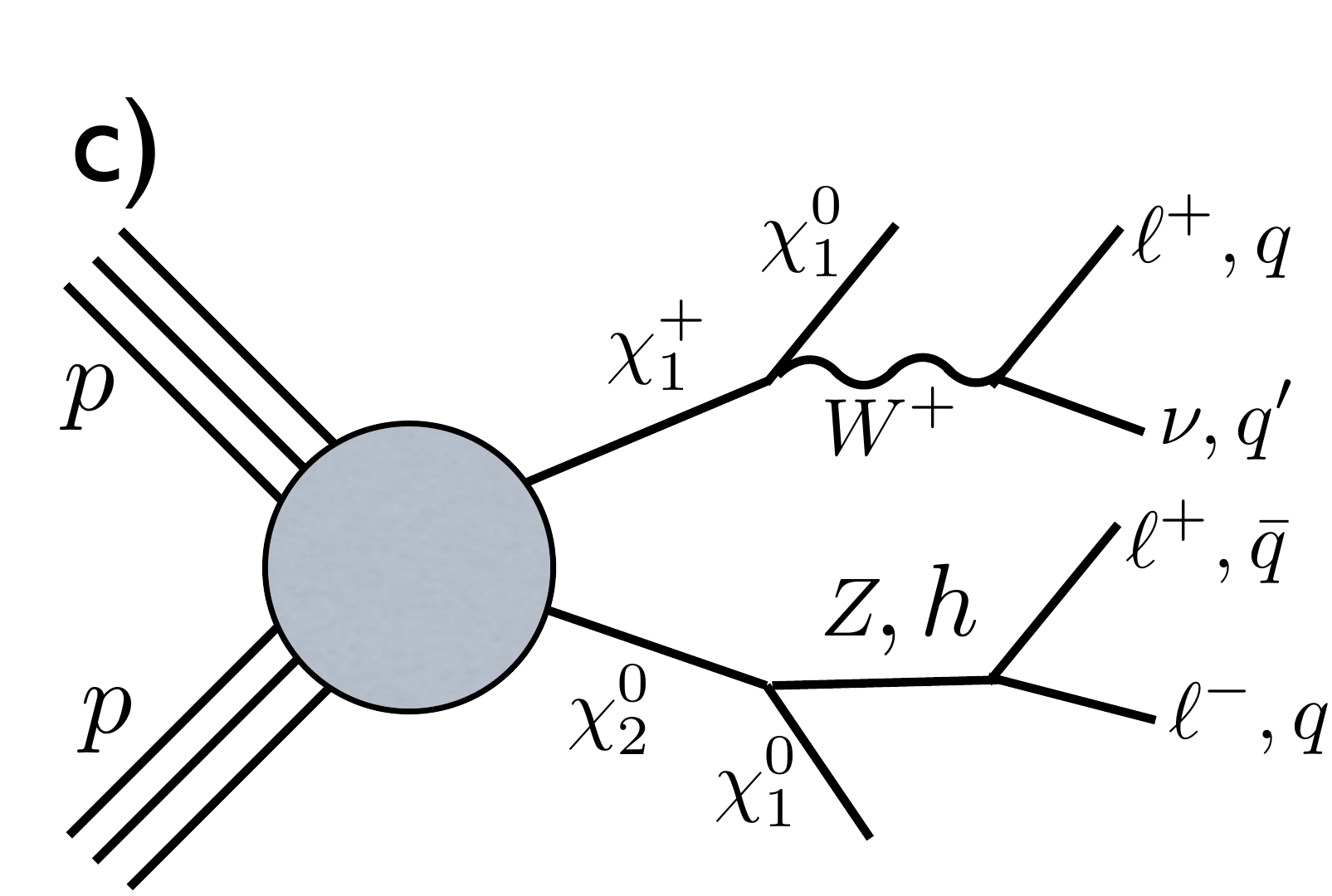}
\includegraphics[width=0.4\textwidth,angle=0]{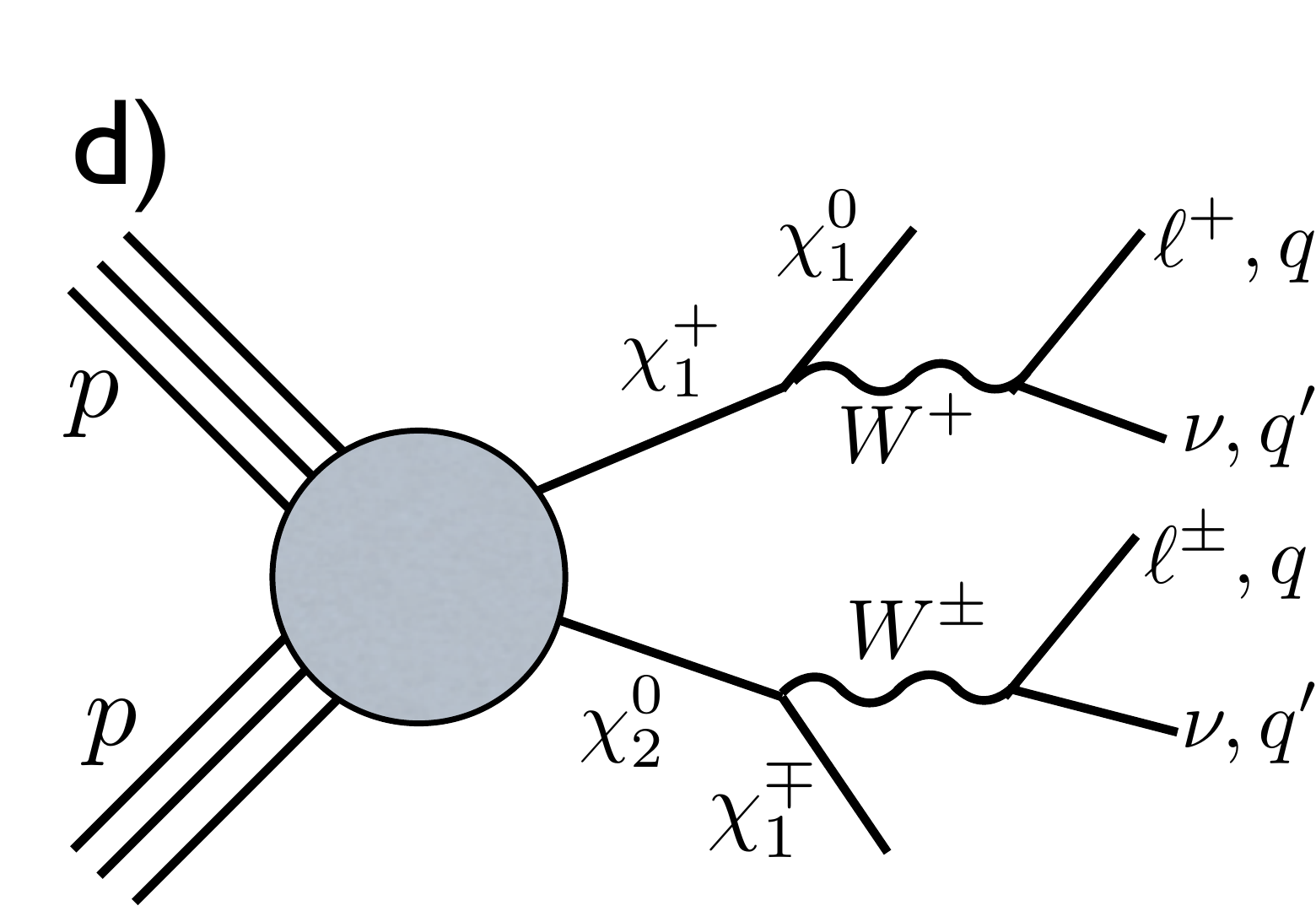}
\caption{Production and decay processes of charginos and neutralinos.}
\label{fig:prod}
\end{figure}

In this section we analyse constraints coming from recent LHC searches for the scenarios discussed in the previous sections, as well as possible future signals. In the first subsection we focus on a possible SUSY-EW contribution to the $WW$ cross section measurement. In subsequent subsections we discuss impact of di- and tri-lepton searches. Finally, we turn to future prospects for discovery of this kind of scenarios at colliders. In all cases the relevant processes arise from chargino and/or neutralino production, whose corresponding diagrams are shown in figure~\ref{fig:prod}. 

In order to find whether our benchmark models are constrained by the published LHC searches we simulate events using \texttt{Herwig++ 2.5.2}~\cite{Bahr:2008pv,Gigg:2007cr} and analyze them using the fast detector simulation package \texttt{Delphes~2.0.3}~\cite{Ovyn:2009tx}. We implement selection procedures and cuts as described in the relevant ATLAS and CMS notes \cite{ATLAS:2012mec,Chatrchyan:2013oev}. This is then validated using cut-flows and efficiencies reported by the collaborations. The signal is scaled to the NLO production rate using \texttt{Prospino~2.1}~\cite{prospino2}. 

From the previous section it is clear that there are two interesting regions to explore: $M_1\gg \mu, M_2$ and some windows at $|M_1|\lesssim 100$ GeV, where $m_{\chi^0_1}\approx m_h/2$. The region of large $M_1$ is neither constrained by the present LHC data nor capable of giving signals in near-future searches. The reason is the following. In that region the lightest chargino is quite degenerate with the lightest neutralino (which is essentially a mixture of wino and Higgsino). Consequently, the dominant production processes are $\chi^+_1 \chi^-_1$ (diagram (a) of figure~\ref{fig:prod}) and $\chi^0_1 \chi^\pm_1$ (diagram (b)). Then the leptons produced by the decay of $\chi^\pm_1$ are extremely soft and escape detection. Production of $\chi^0_2 \chi^\pm_1$ is also possible. Then the neutralino can decay as in diagram (c) or (d) of figure~\ref{fig:prod}. In the first case the visible signal are two opposite-sign leptons plus missing energy, which in principle contribute to $WW$ measurement. However, the contribution is never significant. In the second case, the visible signal is a single lepton, which has a large SM background. On the other hand, as commented in the previous section, this region can be tested in the future by XENON1T, since the direct detection cross-section is not far from the present XENON100 limit, as it is apparent from figures~\ref{fig:WX} and~\ref{fig:WXrest}.

Consequently, we focus next on the windows at $|M_1|\lesssim 100$ GeV. For the sake of concreteness we analyze the windows corresponding to positive $M_1$ and $m_{\chi^0_1}\approx m_h/2$ for the five benchmarks (Models \#1--5). The corresponding masses of charginos and neutralinos, cross sections for $pp\to \chi_1^+\chi_1^-$, $pp\to \chi_1^\pm\chi_2^0$ at $\sqrt{s} = 8$~TeV, and dominant decay modes are listed in table~\ref{tab:masses}.

\begin{table}
\begin{center}
 \begin{tabular}{lrrrrr} \toprule
                  & \#1       & \#2      & \#3      & \#4      & \#5        \\ \toprule
  $m_{\chi_1^0}$  & $62.4$  & $61.3$  & $62.1$  & $61.9$  & $62.1$   \\ \hline
  $m_{\chi_2^0}$  & $125.8$ & $151.0$ & $155.7$ & $148.7$ & $114.7$  \\ \hline
  $m_{\chi_3^0}$  & $185.0$ & $220.0$ & $200.0$ & $250.0$ & $173.6$  \\ \hline
  $m_{\chi_4^0}$  & $270.7$ & $304.7$ & $312.2$ & $314.4$ & $238.8$  \\ \hline
  $m_{\chi_1^\pm}$ & $104.5$ & $139.5$ & $140.7$ & $140.7$ & $104.1$  \\ \hline
  $m_{\chi_2^\pm}$ & $265.5$ & $300.5$ & $309.3$ & $309.3$ & $237.6$  \\ \hline
  $\sigma(\chi_1^+\chi_1^-)$ & $2.7$ & $0.9$  &  $0.7$ & $1.0$ & $2.7$  \\ \hline
  $\sigma(\chi_1^\pm\chi_2^0)$  & $2.6$ & $1.3$  & $1.0$  & $1.7$ & $3.5$ \\ \hline
  $\chi_1^\pm \to \chi_1^0 W^*$      & $100\%$ & $100\%$ & $100\%$ & $100\%$ & $100\%$  \\ \hline 
  $\chi_2^0 \to \chi_1^\pm W^*$      & $100\%$ & $93\%$  &         & $81\%$  &          \\ \hline
  $\chi_2^0 \to \chi_1^0 h^*$        &         & $7\%$   &         & $19\%$  &          \\ \hline
  $\chi_2^0 \to \chi_1^0 Z^{(*)}$    &         &         & $100\%$ &         & $100\%$  \\ \bottomrule
 \end{tabular}
\caption{Masses (in GeV), cross sections (in pb, $\sqrt{s} = 8$~TeV) and dominant decay branching ratios of charginos and neutralinos in five representative models. \label{tab:masses}}
\end{center}
\end{table}

\subsection{$WW$ measurement and SUSY-EW contribution}

The most accurate measurement of the $WW$ production cross section was performed by the ATLAS and CMS experiments in the dilepton plus missing energy channel~\cite{ATLAS:2012mec,Chatrchyan:2013oev,CMS:2012cva}. Both experiments have reported a slight, but statistically not significant, excess. It was suggested that such an excess could have supersymmetric origin, either due to light charginos~\cite{Curtin:2012nn} or stops~\cite{Rolbiecki:2013fia}. It is therefore interesting to see whether the models considered in the present study could also give a contribution in this channel.

The dilepton signal can originate from diagrams (a), (c) and (d) of figure~\ref{fig:prod}. The necessary ingredients leading to a significant contribution are a sizeable cross section and leptons that are hard enough to pass experimental selection criteria. Since our benchmark models are in a strong mixing regime, the couplings involved in the production process tend to be large, which increases the cross section. In addition, the fact that the chargino masses are not far to the LEP bound also increases the cross section. In order to provide high $p_T$ leptons, the mass difference between charginos and neutralinos should be sizable; in our case, close to the $W$ boson mass. On the other hand, for diagram (c), that is typical for Models \#3 and \#5 (see table~\ref{tab:masses}), one of the leptons from $Z^{(*)}$ decay should escape detection in order to contribute to the dilepton final state. In contrast, for diagram (d), which is typical for Models \#1, \#2 and \#4, the lepton coming from the decay of the off-shell $W$ in neutralino ${\chi}_2^0$ is usually missing due to the small mass difference between ${\chi}_2^0$ and ${\chi}_1^\pm$. 

In a detailed simulation we have found that, among our benchmarks, Models \#2 and \#4 give the largest contribution to the $WW$ cross section measurement. At 8~TeV, for the CMS analysis, they increase the total event yield by $\sim 50$ events, which is about half of the observed excess after taking into account the contribution of the Higgs bosons \cite{Chatrchyan:2013oev}. The contribution for the remaining models is below 20 events. It is worth noting that even though na\"{\i}vely one could expect Models \#3 and \#4 to be very similar here, the net result turns out to be very different.  This is caused by the neutralino ${\chi}_2^0$ being slightly heavier in the former case, resulting in a decay via an on-shell $Z$ (see diagram (c) of figure~\ref{fig:prod}) rather than via chargino and an off-shell $W$ (diagram (d) of figure~\ref{fig:prod}), as happens for Model \#4. As can be seen in table~\ref{tab:masses}, for Model \#4 the mass difference $m_{\chi^\pm_1} - m_{\chi_2^0}$ is just 8~GeV, which guarantees that the lepton is soft and would never be detected. Incidentally, this example shows how light gauginos can evade detection due to a multitude of possible decay chains, a fact that is usually missed by simplified models.    

\subsection{Trilepton final states}

Trilepton searches~\cite{ATLAS-CONF-2013-035} directly target SUSY-EW production at the LHC, having at the same time low SM background. The contribution to the trilepton final state comes from ${\chi}_2^0{\chi}_1^\pm$ production and subsequent decay through diagrams (c) and (d) of figure~\ref{fig:prod}. However, as already discussed, in case (d) one lepton from  ${\chi}_2^0$ decay tends to be too soft to pass selection criteria, since ${\chi}_2^0$ and ${\chi}_1^\pm$ are close in mass. Therefore one can expect that this contribution will be usually suppressed. On the other hand, diagram (c) could significantly contribute to the trilepton final state.\footnote{See \cite{Cabrera:2012gf} for a detailed discussion of detection strategies for these kind of processes.}

Out of the five benchmarks, Models \#1 and \#2, \#3, \#4 are not constrained by the ATLAS search \cite{ATLAS-CONF-2013-035}. As expected, Model \#3 gives a non-negligible contribution (recall that in this case the neutralino decays as in diagram (c)), namely $\sim 30$ events at 8 TeV with 20.7 fb$^{-1}$, in the signal region with an on-shell $Z$, but still well below the $95\%$ CL reported by the experiment. It turns out, however, that Model \#5 is excluded by this search. This occurs because the dominant decay mode of the neutralino ${\chi}_2^0$ is via an off-shell $Z$ boson. The combination of a high production cross-section and favourable decay branching ratios leads to a high event yield and, hence, to the exclusion of the model. Incidentally (and unfortunately), the same exclusion occurs for Model \#5 using $M_1\simeq -49.5$ GeV, a choice that, as discussed at the end of section~\ref{sec:4}, led to the observed DM abundance and was allowed by direct-detection limits. This is a general conclusion for models with negative $M_1$ that reproduce the correct DM abundance. As can be seen in figure~\ref{fig:masses}, for $M_1 \simeq 50$-$60$~GeV the second neutralino and the light chargino are almost mass degenerate. Then, due to kinematical reasons the $\chi_2^0 \to \chi_1^0 Z^{(*)}$ decay is strongly preferred  to the $\chi_2^0 \to \chi_1^\pm W^{*}$ one, hence yielding high trilepton signal rate.

\subsection{Same-sign dilepton final states}

Dilepton final states typically result from the production of sleptons and charginos. Contrary to the trilepton searches, SM backgrounds are much larger in this case and, hence, more tight selection criteria are required~\cite{Aad:2012pxa}. One exception to this is a final state with two same-sign leptons (SS), which has a low SM background, mostly due to fake leptons, charge mismeasurement and diboson production. In our case, the genuine SS-leptons signal originate from the diagram (d) of figure~\ref{fig:prod}.\footnote{The other possibility arises when one of the leptons in diagram (c) is missing, hence making the dilepton search complementary to the trilepton searches. } This decay mode  of ${\chi}_2^0$ is dominant for Models \#1, \#2 and \#4. Indeed, Models \#2 and \#4 give the largest contribution here, but still below the experimental exclusion limits. None of our benchmark models is therefore excluded by SS dilepton searches.

\subsection{Future prospects}

\begin{figure}[t!]
\centering
\includegraphics[width=0.49\textwidth,angle=0]{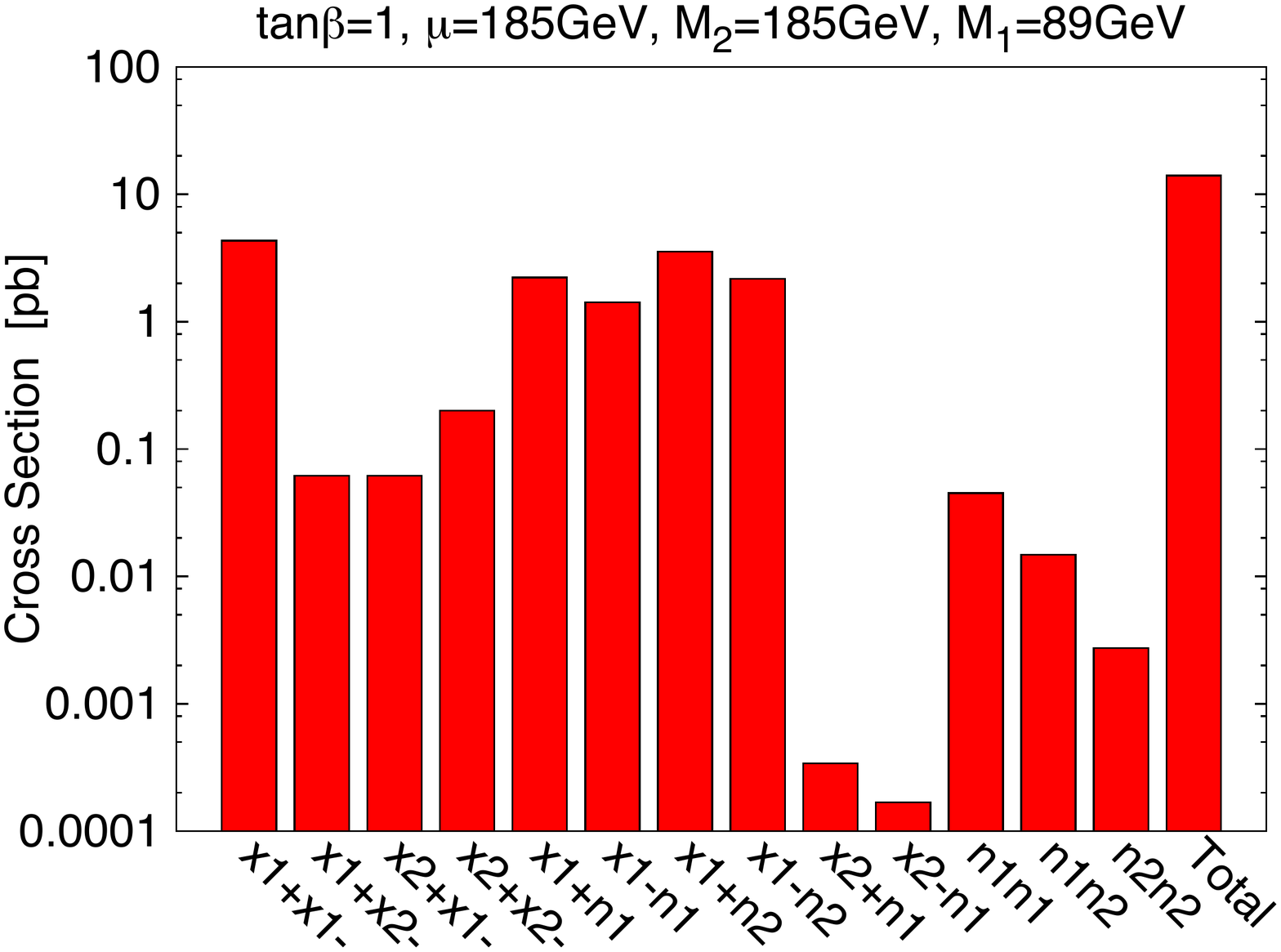}
\includegraphics[width=0.49\textwidth,angle=0]{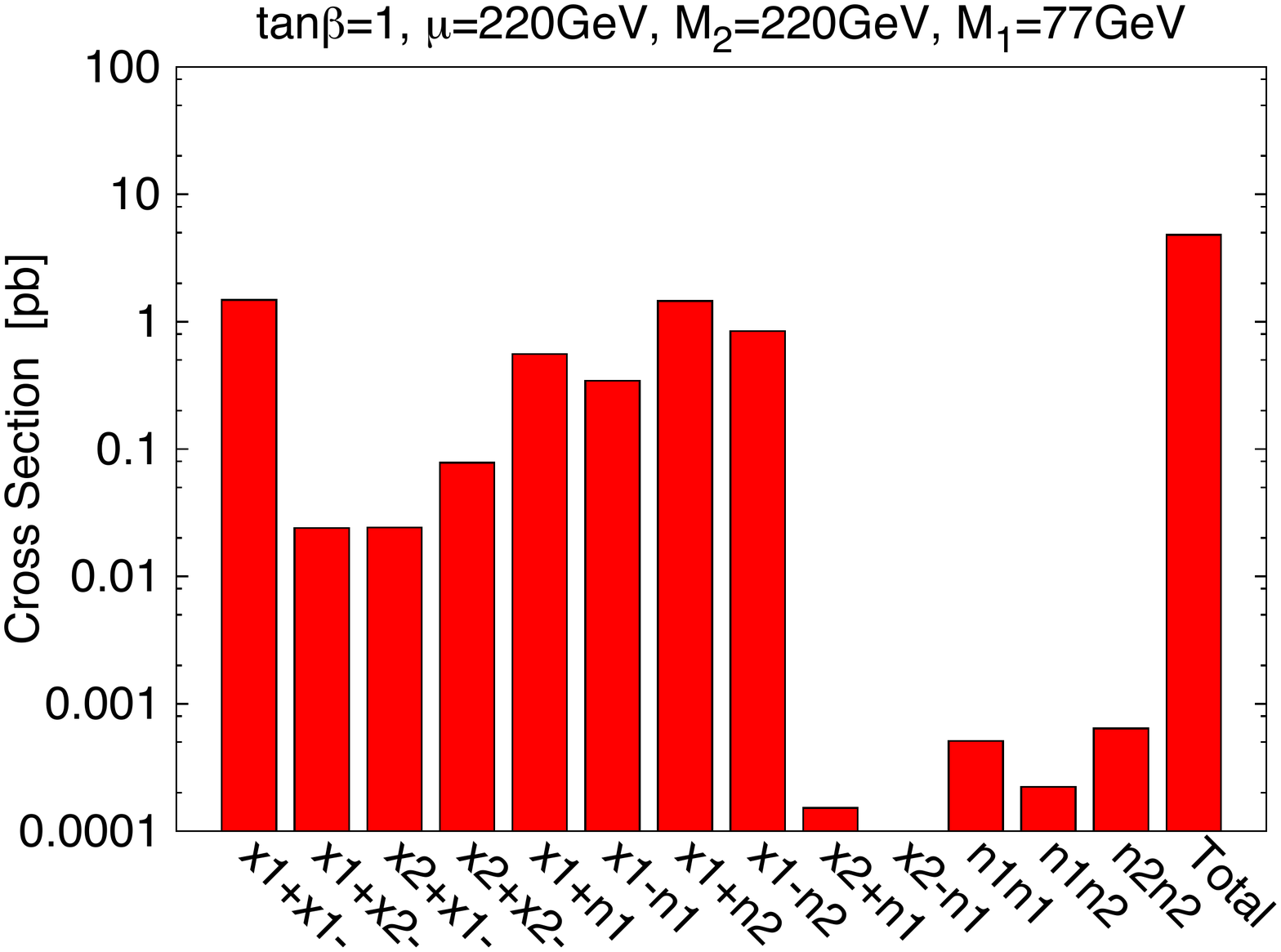}
\includegraphics[width=0.49\textwidth,angle=0]{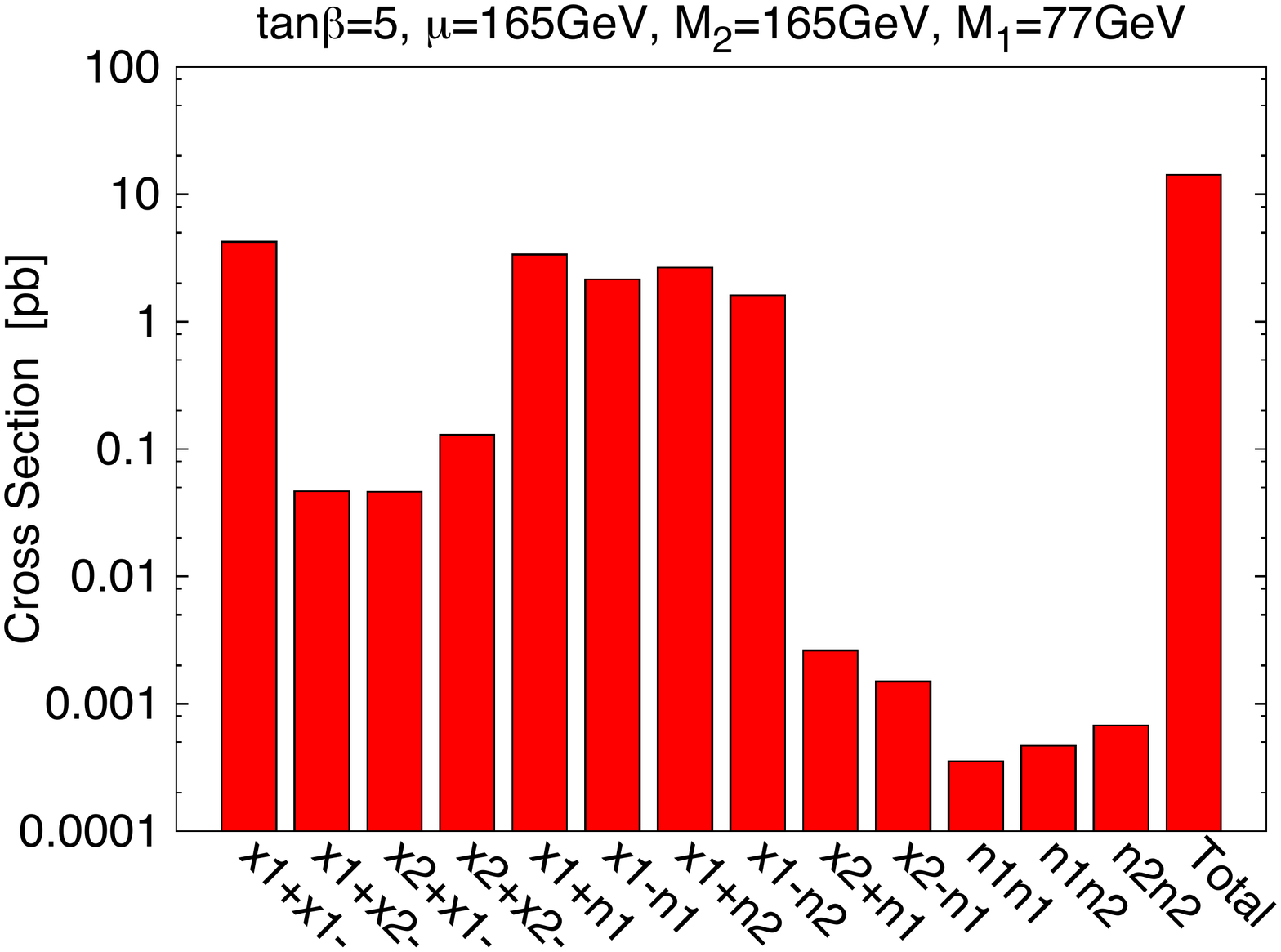}
\caption{{ Cross-sections of chargino pair, neutralino pair and chargino+neutralino production for the different models described in the text.}}
\label{fig:xsec}
\end{figure}

Finally, let us discuss the future prospects of discovery of our benchmark Models \#1, \#2, \#3 and \#4 (recall that Model \#5 was excluded by trilepton searches). The cross sections for chargino and neutralino production at the LHC with $\sqrt{s} = 14$~TeV are summarised in figure~\ref{fig:xsec}.

If the excess in the $WW$ cross section measurement is established with a higher significance, its beyond-SM origin can be probed using a method described in ref.~\cite{Rolbiecki:2013fia}. In such a case, light charginos and neutralinos can be a good candidate to explain the excess. Its electroweakino origin could be eventually confirmed using the SS leptons search for models similar to our benchmark scenarios \#2 and \#4. Based on our simulation, about $\mathcal{L} \sim 100~\mathrm{fb}^{-1}$ would be required for a $5\sigma$ excess in the SS lepton signal region at $\sqrt{s} = 14$~TeV. On the other hand, Models \#1 and \#3 can be expected to show up in the trilepton search.

In any case, a precise measurement of chargino and neutralino properties can only be performed at a planned linear collider. These results could be confronted with other measurements: $h\to \gamma \gamma$, DM abundance and DM direct searches. Such a consistency check could help to establish a new physics model and possibly direct future searches.

\section{Conclusions\label{sec:conclusions}}

The observed mass of the Higgs boson, $m_h\simeq 126$ GeV, has deep implications for the MSSM: namely the stop masses (at least one of them) should be well above 1 TeV, unless the stop mixing is close to the maximal value. This size is larger than expected from naturalness arguments and suggests that the supersymmetric spectrum might live at scales inaccessible to the LHC, both for the direct production/detection of the supersymmetric states and for the measurement of indirect effects on low-energy observables, such as the Higgs decay rates in different channels.

However, it may well occur that a part of the supersymmetric spectrum is light enough to show both direct and indirect measurable effects. An attractive possibility in this sense is that charginos and neutralinos are substantially lighter than sfermions. This scenario is supported not only by the phenomenological fact that the present bounds on charginos and neutralinos are pretty mild, but also by some theoretical hints, such as the successful supersymmetric unification of the gauge couplings (which requires light supersymmetric fermions) and by the appealing possibility that the dark matter is made of neutralinos. Besides, this framework is welcome to avoid flavour violation problems. 

Among the indirect implications of the existence of light charginos and neutralinos, probably the most remarkable ones concern the modification of the Higgs properties; in particular there is a potentially important contribution of charginos to the decay rate into two photons, $h\rightarrow \gamma \gamma$. This is certainly one of the most relevant Higgs observables, although the experimental situation is not yet clear. At the moment ATLAS continues to observe a 2$\sigma$ excess over the SM prediction, while CMS has become consistent with the SM within 1$\sigma$. Combining the two measurements gives a $\sim 20\%$ excess. If that persists in future with reduced uncertainty, the chargino contribution might become a very natural way to explain it without spoiling the other decay rates. In this paper we have re-visited this possibility, examining the implications of an chargino-driven $h\to\gamma\gamma$ enhancement for other observables, such as DM constraints, electroweak observables (namely $S,T,U$ parameters) and experimental signals at the LHC. 

An important feature for this issue is the well-known fact that the chargino contribution to $h\rightarrow \gamma \gamma$ increases for low $\tan\beta$, thus the depicted scenario naturally requires $\tan\beta\simlt 5$. Interestingly, the possibility of low $\tan\beta$ has received much attention in recent times, e.g.\ it becomes natural when supersymmetric particles are heavy~\cite{Ibanez:2013gf} and shows a rich phenomenology~\cite{Djouadi:2013vqa}. Besides $\tan\beta$, the other relevant parameters are the bino and wino masses, $M_1$ and $M_2$, and the $\mu$ parameter. We have explored the regions of this parameter space able to produce a measurable enhancement in $h\rightarrow \gamma \gamma$, and investigated to which extent they are compatible with the observed DM abundance and the latest bounds on direct detection of DM. For the surviving scenarios we have studied the phenomenology in colliders using 5 representative benchmark models. 

The most important constraints and chances of future detection come from chargino-neutralino ($\chi_2^0\chi_1^\pm$) and chargino-chargino ($\chi_1^+\chi_1^-$) production. The cross sections are typically rather high, thanks partially to the low masses but also to the fact that the models are in a strong-mixing regime. Such production gives potential signals for a number of processes: $WW$ measurement (i.e.\ two leptons + missing energy), trilepton final states, same-sign dilepton final states. Indeed, some models give a substantial contribution to the $WW$ measurement, which could be responsible (at least partially) for the moderate excess observed over the SM prediction. On the other hand, one model becomes excluded by the recent ATLAS trilepton search. However, the other models survive easily, in spite of having quite light charginos. Actually, these examples show how light gauginos can evade detection due to a multitude of possible decay chains, which is usually missed by simplified models. Concerning future detection, we have shown that all benchmarks could be detected at $\mathcal{L}\sim 100 \mathrm{fb}^{-1}$ and $\sqrt{s} = 14$~TeV, or (for regions with large $M_1$) by XENON1T or similar experiments of direct DM detection.

In conclusion, models with charginos and neutralinos much lighter than sfermions are interesting both from theoretical and experimental perspectives. If charginos are responsible for the possible $h\rightarrow \gamma \gamma$ excess, this strongly determines the theoretical framework (which requires low to moderate $\tan\beta$ and light charginos), and the available signals at the LHC and in experiments of direct detection of DM.

\acknowledgments
This work has been partially supported by the MICINN, Spain, under contract FPA2010-17747; Consolider-Ingenio  CPAN CSD2007-00042, as well as MULTIDARK CSD2009-00064. We thank as well the Comunidad de Madrid through Proyecto HEPHACOS S2009/ESP-1473 and the European Commission under contract PITN-GA-2009-237920. We also thank the spanish MINECO Centro de excelencia Severo Ochoa Program under grant SEV-2012-0249.

\bibliographystyle{JHEP}
\bibliography{cmrz}

\end{document}